\documentclass[twocolumn,aps,prc,superscriptaddress,showpacs,nobibnotes]{revtex4}
\usepackage{epsfig,dcolumn,bm}

\newcommand{\dmu}{\partial_\mu}

\newcommand{\lc}{{\cal L}}

\newcommand{\lm}{{\cal M}}

\newcommand{\sqd}{\sqrt{2}}

\newcommand{\be}{\begin{eqnarray}}
\newcommand{\ee}{\end{eqnarray}}
\newcommand{\nn}{\nonumber}

\begin{document}

\title{Isospin breaking effects in the $X(3872)$ resonance}

\author{D. Gamermann}{\thanks{E-mail: daniel.gamermann@ific.uv.es}
\author{E. Oset}{\thanks{E-mail: oset@ific.uv.es}
\affiliation{Departamento de F\'isica Te\'orica and IFIC, Centro Mixto
Universidad de Valencia-CSIC,\\ Institutos de Investigaci\'on de
Paterna, Aptdo. 22085, 46071, Valencia, Spain}

\begin{abstract}
In this paper we study the effects of isospin breaking in the dynamical generation of the $X(3872)$ state. We also calculate the ratio of the branching fractions of the $X$ decaying into $J/\psi$ with two and three pions, which has been measured experimentally to be close to unity. Together with the $X(3872)$, of positive C-parity, we predict the existence of a negative C-parity state and we comment on which decay channel is more promising to observe this state. The simultaneous investigation of the $X(3872)$ decay into $J/\psi\pi\pi$ and $D^0\bar D^{*0}$ show a preference for a slightly unbound, virtual state of $D\bar D^*$ and $\bar D D^*$
\end{abstract}


\maketitle


\section{Introduction}

The $X(3872)$ was discovered at Belle \cite{belledisc} and then later was also 
observed at CDFII and D0 collaborations and BaBar \cite{conf1,conf2,conf3}.
In all these experiments the $X$ has been discovered and observed in the decay channel
$J/\psi\pi^+\pi^-$. There is strong evidence that the dipion generated in this decay
channel comes from a $\rho$ meson \cite{cdf2pi}. Later on also the decays of the $X$ into 
$J/\psi\pi^+\pi^-\pi^0$ and $J/\psi\gamma$ have been observed \cite{bellegj}, this latter
decay channel indicating that the C-parity of the $X$ is positive. The quantum numbers of the $X(3872)$
have been investigated in \cite{xqn}, concluding that it must
correspond to $J^P=1^{++}$ or $J^P=2^{-+}$. Observed only on a neutral charge
state it is assumed to have isospin $I=0$.  Its decay into   $J/\psi \eta$ has
been investigated in \cite{nojeta} but only an upper bound has been
found. The non observation of the decay $J/\psi\eta$ is a further evidence of the positive C-parity of the $X$.
 It could also mean that in the particular
reaction of \cite{nojeta} the $X(3872)$ was necessarily produced with
positive $C$-parity, without ruling out the possibility of a nearby state with
negative $C$-parity. The existence of two nearly degenerate $X(3872)$ states
appears in some theoretical models \cite{Terasaki:2007uv,danielaxial}. The most
popular view about the nature of this resonance is that it is made of $D
\bar{D}^*$ \cite{danielaxial,Liu:2008fh,Liu:2007bf,Dong:2008gb,Swanson:2003tb,gutsch1,gutsch2}, a
recent review can be seen in \cite{Liu:2008du}. One of the problems faced by
these models is the large ratio for

\be
\frac{{\cal B}(X\rightarrow J/\psi\pi^+\pi^-\pi^0)}{{\cal B}(X\rightarrow J/\psi\pi^+\pi^-)}&=&1.0\pm0.4\pm0.3\textrm{ .}
\ee
Indeed, since the resonance has positive C-parity the denominator can go via $J/\psi\rho$ as
supported by the experiment \cite{cdf2pi}. However, the $X(3872)$ state has $I=0$ and 
then isospin is violated. On the contrary the numerator can go through 
$J/\psi\omega$ as supported by experiment \cite{bellegj}, in which case there is no
violation of isospin. The fact that the ratio is so large in spite of the violation of
isospin found a plausible explanation in \cite{Swanson:2003tb,gutsche}, where the state
was supposed to be largely $D^0 \bar{D}^{*0}$ but with some coupling to both 
$J/\psi\omega$ and $J/\psi\rho$. Even if the coupling to $J/\psi\rho$ is
small, as expected from isospin symmetry breaking, the larger phase space for 
$J/\psi\rho$ decay than for $J/\psi\omega$, because of the large width of the
$\rho$, can account for the large ratio.  Although other  charged 
$D \bar{D}^*$ components can appear in the wave function, the neutral charge
component is preferred since it is the one closest to threshold and hence
should have the largest weight. The idea is intuitive and widely accepted, see
\cite{braaten}.  The idea on the dominance of the neutral component is worth
pursuing. Indeed. In \cite{danielaxial}, where a dynamical theory for the
generation of the $X(3872)$ resonance based on the hidden gauge approach for the
vector-meson interaction was done, isospin symmetry was kept and the masses of
the charged and neutral $D$ mesons were taken equal. The fact that the binding
energy for the $D^0 \bar{D}^{*0}$ is so small advises to revise the model to
account for the mass differences with the charged $D \bar{D}^*$, which can induce isospin breaking and a
dominance of the  $D^0 \bar{D}^{*0}$ in the wave function. In the present paper
we will face this problem and will discuss qualitatively as well as quantitatively,
the limit of zero binding energy, with interesting results. We will also revisit
the interpretation of the reaction production of $D\bar D^*$ \cite{expcross}
showing that it gives support to the existence of the $X(3872)$ as a 
$D \bar{D}^*$ narrow state of positive parity, without ruling out the possible
existence of a broader one of negative parity. We discuss the decay modes of the
negative C-parity state and speculate on where could it be found and the
possible difficulties in its observation.

The present work should also be looked at with some perspective. Although most of the work
for the $X(3872)$ has concentrated on the molecular $D\bar D^*$ picture, as we have mentioned,
since the mass of is so close to the threshold, there are other pictures proposed to describe it
in terms of quarks, tetraquarks and other dynamical pictures (see \cite{qq1,qq2,qq3} for reviews).
The molecular picture of some meson resonances is catching up, particularly when it comes to interpret
states that do not fit clearly in a $q\bar q$ picture. This is the case of the $X$, $Y$ and $Z$ resonances
recently discovered at the $B$-factories for which there are also other structures proposed and about which
there is an intensive debate (see \cite{olsen1} for a recent overview on the subject).

This work is organized as follows: in the next section we shortly explain our phenomenological
Lagrangian and present our framework to generate dynamically resonances from the interaction
of pseudoscalars with vector-mesons. In the same section we show results with and without isospin
violation and we also show what happens when the binding energy goes to zero. In section III we
comment on the two C-parity states that our model generates and in section IV we make our
final remarks and conclusions.


\section{Theoretical Model}

The framework that we describe here is explained in more details in \cite{danielaxial} and references therein.

We want to study the interaction of a pseudoscalar with a vector-meson that, in s-wave, has the quantum numbers 
of an axial: $1^+$.
The starting point of our model consists of the fields belonging to the 15-plet and a singlet of $SU(4)$ describing the 
pseudoscalar and vector-mesons:

\begin{widetext}
\be
\Phi&=&\left(
\begin{array}{cccc}
 \frac{\eta }{\sqrt{3}}+\frac{\pi^0}{\sqrt{2}}+\frac{\eta'
   }{\sqrt{6}} & \pi ^+ & K^+ & \overline{D}^0 \\& & & \\
 \pi ^- & \frac{\eta }{\sqrt{3}}-\frac{\pi
   ^0}{\sqrt{2}}+\frac{\eta'}{\sqrt{6}} & K^0 & D^- \\& & & \\
 K^- & \overline{K}^0 & \sqrt{\frac{2}{3}} \eta'-\frac{\eta
   }{\sqrt{3}} &  {D_s}^- \\& & & \\
 D^0 & D^+ &  {D_s}^+ & \eta _c
\end{array}
\right) \\
\cal{V}_\mu&=&\left( \begin{array}{cccc}
{\rho_\mu^0 \over \sqd}+{\omega_\mu \over \sqd} & \rho^+_\mu & K^{*+}_\mu & \bar D^{*0}_\mu \\ & & & \\
\rho^{*-}_\mu & {-\rho^0_\mu \over \sqd}+{\omega_\mu \over \sqd} & K^{*0}_\mu & D^{*-}_\mu \\& & & \\
K^{*-}_\mu & \bar K^{*0}_\mu & \phi_\mu & D_{s\mu}^{*-} \\& & & \\
D^{*0}_\mu & D^{*+}_\mu & D_{s\mu}^{*+} & J/\psi_\mu \\ \end{array} \right).
\ee
\end{widetext}
Note that these fields differ from those used in \cite{danielaxial} because of the inclusion of $\eta$-$\eta'$
and $\omega$-$\phi$ mixing.

For each one of these fields a current is defined:

\be
J_\mu&=&(\dmu \Phi)\Phi-\Phi\dmu\Phi \\
\cal{J}_\mu&=&(\dmu \cal{V}_\nu)\cal{V}^\nu-\cal{V}_\nu\dmu \cal{V}^\nu. 
\ee

The Lagrangian is constructed by coupling these currents:

\be
\lc_{PPVV}&=&-{1\over 4f^2}Tr\left(J_\mu\cal{J}^\mu\right). \label{lag}
\ee

In the way it is constructed this Lagrangian is $SU(4)$ symmetric, but we know that $SU(4)$ symmetry is badly
broken in nature. To take this into account we will break the $SU(4)$ symmetry of the Lagrangian in the following way:
Assuming vector-meson dominance we recognize that the interaction behind our Lagrangian is the exchange of a vector
meson in between the two hadronic currents. If the initial and final pseudoscalars (and vector-mesons), in a given process,
have different charm quantum number, it means that the vector-meson exchanged in such a process is a charmed meson,
and hence a heavy one. In these cases we suppress the term in the Lagrangian containing such processes by a factor $\gamma=m_L^2/m_H^2$ where $m_L$ is the typical value of a light vector-meson mass (800 MeV) and $m_H$ the typical value of the heavy vector-meson mass (2050 MeV). We also 
suppress, in the interaction of $D$-mesons the amount of the interaction which is driven by a $J/\psi$ exchange by the factor
$\psi=m_L^2/m_{J/\psi}^2$. Another source of symmetry breaking will be the meson decay constant $f$ appearing in the Lagrangian.
For light mesons we use $f=f_\pi=93$ MeV but for heavy ones $f=f_D=165$ MeV.

So, for a given process $(P(p)V(k))_i\rightarrow (P'(p')V'(k'))_j$ we have the amplitude:

\be
\lm_{ij}(s,t,u)&=&-{\xi_{ij}\over4f_i f_j}(s-u)\epsilon . \epsilon ' \label{ampli}
\ee
where $s$ and $u$ are the usual Mandelstam variables, $f_i$ is the pseudoscalar $i$ meson decay constant, $\epsilon$ are the vector-meson polarization vectors and $i$, $j$ refer to the initial and final channels in the coupled channel space. The coefficient matrices $\xi_{ij}$ can be directed calculated from the Lagrangian of eq. (\ref{lag}) in charge basis. The $\xi_{ij}$ coefficients in charge basis are given in the appendix.

The amplitude in eq. (\ref{ampli}) is projected in s-wave and plugged into the scattering equation for the coupled channels:

\be
T&=&V+VGT. \label{bseq}
\ee
In this equation $G$ is a diagonal matrix with each one of its elements given by the loop function for each channel in the coupled channel space. For channel $i$ with mesons of masses $m_1$ and $m_2$ $G_{ii}$ is given by:

\be
G_{ii}&=&{1 \over 16\pi ^2}\biggr( \alpha _i+Log{m_1^2 \over \mu ^2}+{m_2^2-m_1^2+s\over 2s}
  Log{m_2^2 \over m_1^2}\nonumber\\ 
 &+ &{p\over \sqrt{s}}\Big( Log{s-m_2^2+m_1^2+2p\sqrt{s} \over -s+m_2^2-m_1^2+
  2p\sqrt{s}}\nn\\
&+&Log{s+m_2^2-m_1^2+2p\sqrt{s} \over -s-m_2^2+m_1^2+  2p\sqrt{s}}\Big)\biggr) \label{loop}
\ee
where $p$ is the three momentum of the two mesons in the center of mass frame. The two parameters $\mu$ and $\alpha$ are not independent, we fix $\mu$=1500 MeV and change $\alpha$ to fit our results within reasonable values in the natural range \cite{hyodo}. We actually use two $\alpha$ as free parameters, one for loops with only light mesons in the channel $i$, we call $\alpha_L$ and set it to $\alpha_L$=-0.8, the value used in \cite{danielaxial} to fit the low lying axial resonances. For loops with heavy particles we use $\alpha_H$ and we comment, in what follows, the effects of changing this parameter. Note that just the combination $\alpha-Log(\mu^2)$ is the free parameter in $G$ of eq. (\ref{loop}). The parameter $\mu$ is there to set a scale of energies, it can be chosen arbitrarily and hence is not related with the cut off in three momentum which can be alternatively used to evaluate $G_{ii}$ from the loop function of two meson propagators. Once $\mu$ is fixed then there is a relationship between $\alpha$ of eq. (\ref{loop}) and an equivalent cut off \cite{ollerloop}. In any case, the equivalence of the cut off method and the dimensional regularization of eq. (\ref{loop}) in a certain region of energies requires the cut off to be reasonably bigger than the on shell three momenta of the intermediate states, which is fulfilled in the energy regime that we study here. In the present case, the $D\bar D^*$ states, with 20 MeV above threshold have a three momentum around 195 MeV, while the cut offs are of the order of 650-850 MeV, as we shall see later on.

The imaginary part of the loop function ensures that the T-matrix is unitary, and since this imaginary part is known, it is possible to do an analytic continuation for going from the first Riemann sheet to the second one. Possible physical states (resonances) are identified as poles in the T-matrix calculated in the second Riemann sheet for the channels which have the threshold below the resonance mass.


\subsection{Isospin Symmetric Case}

If we set the masses of all mesons belonging to a same isospin multiplet to a common value, our results will be isospin symmetric. 
Moreover we can consider the transformation under C-parity of pseudoscalar and vector-mesons in order to construct C-parity
symmetric states:

\be
\hat{C} P &=&\bar P \\
\hat{C} V &=&-\bar V
\ee

States like $D\bar D^*$ and $\bar D D^*$ mix up to form a positive and a negative C-parity state. The same happens for the kaons and the $D_s$ mesons. If one writes the $\xi_{ij}$ coefficients that appear in the amplitude of eq. (\ref{ampli}) in C-parity basis for all two meson states with quantum numbers C=0 and S=0, the coupled channel space splits into two, a positive and a negative C-parity part. These coefficients are given in the appendix.

In charge basis, for positive C-parity one has the following channels: $\bar K^{*0}K^0-c.c.$, $\rho^+\pi^--c.c.$, $\bar D^{*0}D^0-c.c.$,  $D^{*+}D^--c.c.$, $D_s^{*+}D_s^--c.c.$ and $K^{*+}K^--c.c.$. While for negative C-parity the channels are: $\rho^+\pi^-+c.c.$, $K^{*+}K^-+c.c.$, $\rho^0\pi^0$, $\omega\pi^0$, $\phi\pi^0$, $\rho^0\eta$, $\rho^0\eta\prime$, $\bar K^{*0}K^0+c.c.$, $D^{*+}D^-+c.c.$, $\bar D^{*0}D^0+c.c.$, $\rho^0\eta_c$, $J/\psi\pi^0$, $\omega\eta$, $\phi\eta$, $\omega\eta\prime$, $\phi\eta\prime$, $\omega\eta_c$, $\phi\eta_c$, $J/\psi\eta$, $J/\psi\eta\prime$, $D_s^{*+}D_s^-+c.c.$ and $J/\psi\eta_c$.

For the masses of the mesons we use the following values:

$m_\pi$=137.5 MeV, $m_K$=496 MeV, $m_\eta$=548 MeV, $m_D$=1867.5 MeV, $m_{D_s}$=1968 MeV, $m_{\eta_c}$=2980 MeV, $m_{\eta\prime}$=958 MeV, $m_\rho$=775 MeV, $m_{K^*}$=894 MeV, $m_\omega$=783 MeV, $m_\phi$=1019 MeV, $m_{D^*}$=2008.5 MeV, $m_{D_s^*}$=2112 MeV and $m_{J/\psi}$=3097 MeV.

If we set $\alpha_H$=-1.34, which is equivalent to a cut-off of 830 MeV in the three momentum, we get two poles with opposite C-parity, the positive one at 3866 MeV with a width smaller than 1 MeV and the negative one at (3875-25$i$) MeV, which means a width around 50 MeV. The poles appear in isospin I=0, as we determine from combining the charge states into definite isospin states. Now while increasing the value of $\alpha_H$ (lowering the cut-off) the poles approach the threshold (at 3876 MeV in the isospin symmetric case). The negative C-parity pole touches the threshold for $\alpha_H$ values bigger that -1.33 (cut-off of 820 MeV), while the positive C-parity one reaches the threshold for $\alpha_H$ around -1.185 (cut-off equivalent to 660 MeV). Once the pole crosses the threshold it does not appear in the second Riemann sheet, it is no longer a resonance, but becomes a virtual state. Yet a peak can be seen in the cross section of some channels, but can not be identified as a pole in the second Riemann sheet of the T-matrix.


\subsection{Isospin Breaking}

In order to investigate the isospin breaking we define the following quantities:

$\Delta m_\pi$=2.5 MeV, $\Delta m_K$=-2 MeV, $\Delta m_D$=2.5 MeV, $\Delta m_{K^*}$=-2 MeV and $\Delta m_{D^*}$=1.5 MeV. In this way the masses of the members of a multiplet split: for the charged members of a multiplet the mass will be equal to $m+\Delta m$ while for the neutral members it will be $m-\Delta m$.

Now there are two $\bar D D^*$ thresholds nearby, the neutral one at 3872 MeV and the charged one at 3880 MeV. The $X(3872)$ state is a very weakly $D^0 \bar D^{*0}$ bound state and the fact that the binding energy is much smaller than the difference between these two thresholds could reflect itself in a large isospin violation in observables.

For simplicity let us consider, for the moment, a toy model with only two channels, with neutral and charged $D$ and $D^*$ mesons. In this model we assume the potential $V$ to be a 2x2 matrix:

\be
V&=&\left(\begin{tabular}{cc} $v$ & $v$ \\ $v$ &$v$ \end{tabular}\right),
\ee
with $v$ constant, which indeed is very close to the real one in a small range of energies.

In this case the solution of the scattering equation (\ref{bseq}) is:

\be
T&=&\frac{V}{1-vG_{11}-vG_{22}}
\ee
where $G_{11}$ and $G_{22}$ are the loop function calculated for channels 1 and 2 respectively. If there is a pole at $s$=$s_R$ we can expand $T$ close to this pole as:

\be
T_{ij}&=&\frac{g_ig_j}{s-s_R}
\ee
where $g_i$ is the coupling of the pole to the channel $i$. The product $g_ig_j$ is the residue at the pole and can be calculated with:

\be
\lim_{s\rightarrow s_R} (s-s_R)T_{ij}&=&\lim_{s\rightarrow s_R} (s-s_R) \frac{V_{ij}}{1-vG_{11}-vG_{22}}
\ee
We can apply the l'H\^opital rule to this expression and we get:

\be
\lim_{s\rightarrow s_R} (s-s_R)T_{ij}&=& \frac{V_{ij}}{-v( \frac{dG_{11}}{ds}+\frac{dG_{22}}{ds})} \label{lhopt}
\ee

If one has a resonance lying right at the threshold of channel 1 the couplings $g_i$ will be zero, since the derivative of the loop function $G_{11}$, in the denominator of eq. (\ref{lhopt}) is infinity at threshold. This is a general property which has its roots in basic Quantum Mechanics as shown in \cite{jnieves}. In figure \ref{fig1} we show plots of the real part of the loop function for the neutral and charged $D$ meson channels.

It is interesting to note that eq. (\ref{lhopt}) for just one channel is the method used to get couplings of bound states to their building blocks in studies \cite{gutsche,hana0} of dynamically generated states following the method of the compositness condition of Weinberg \cite{weinberg,efimov}.

We will come back to this issue again with the realistic model. Now, in what follows, the arguments used do not require the toy model any longer.

\begin{figure}
\begin{center}
\includegraphics[width=5cm,angle=-90]{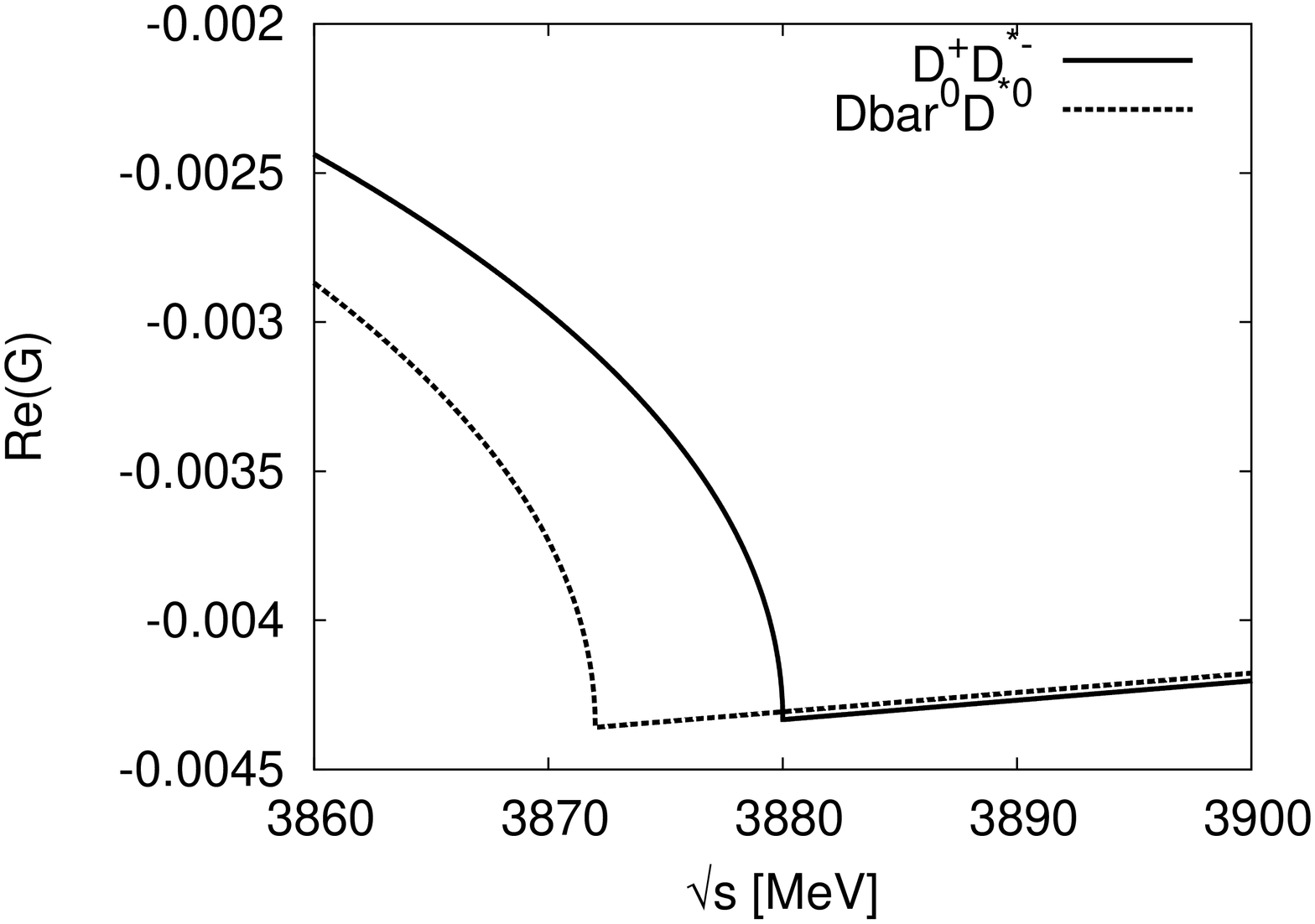}
\caption{Loops} \label{fig1}
\end{center}
\end{figure}

Suppose that the $X(3872)$ decays through the diagram in figure \ref{fig2}. In this figure the $D$ mesons can be either charged or neutral. For the isospin I=1 state with the $\rho$ meson in the final state, the diagrams with neutral $D$ mesons interfere destructively with those with charged $D$ mesons, while in the $\omega$ case they sum up. If the vertices have the same strength for $\rho$ and $\omega$ production (this is the case in the framework of the hidden gauge formalism \cite{hidden1,hidden2,raquelhidden}) the ratio of the amplitudes will be given by the ratio of the difference between the charged and neutral loops divided by the sum of the loops:

\be
R_{\rho/\omega}&=&\left(\frac{G_{11}-G_{22}}{{G_{11}+G_{22}}}\right)^2 \label{ratio}
\ee

\begin{figure}[h]
\begin{center}
\includegraphics[width=5cm,angle=-0]{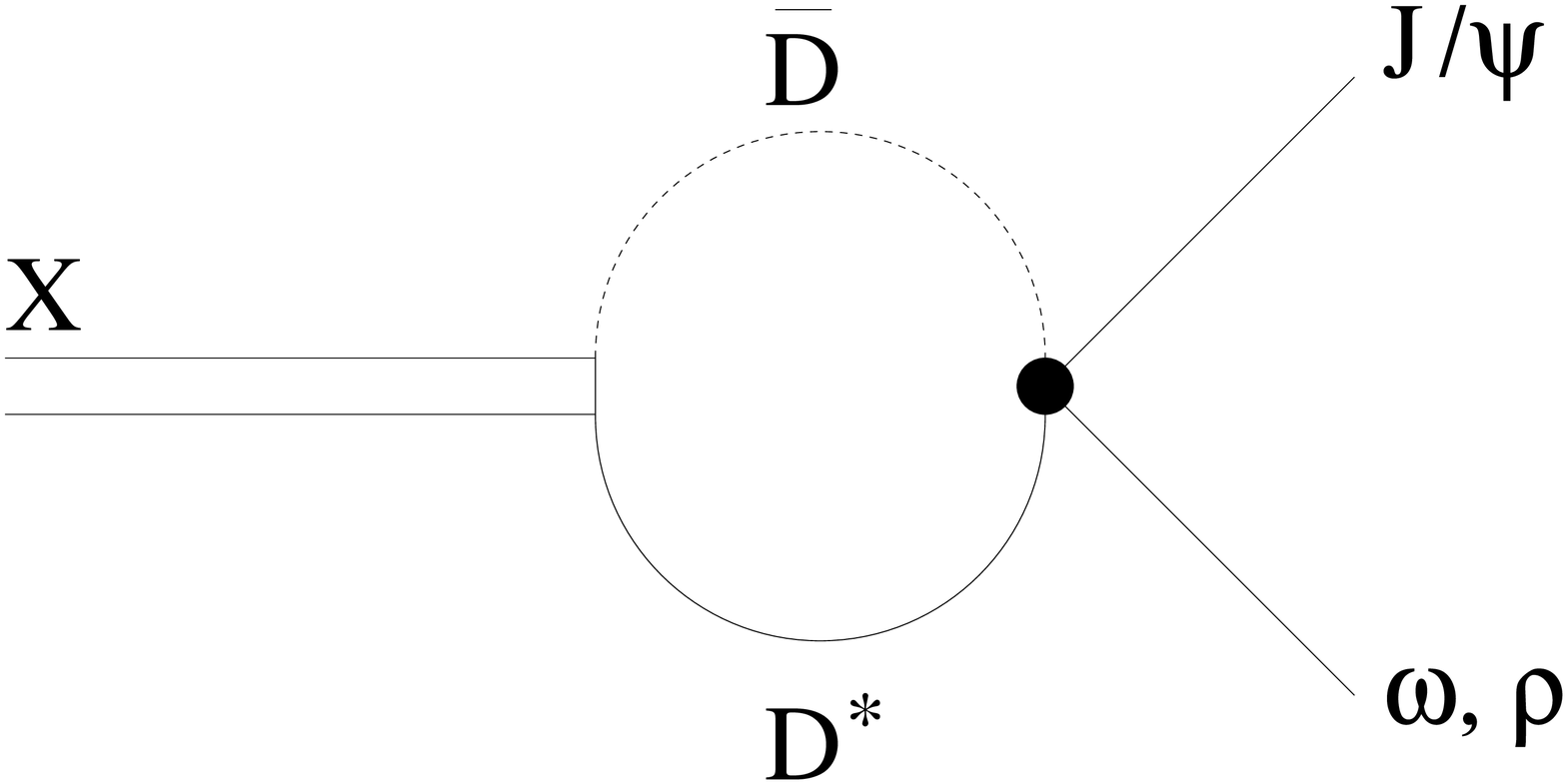}
\caption{$X$ decay} \label{fig2}
\end{center}
\end{figure}

In the isospin symmetric case, the charged and neutral loops are equal, because these loops depend only on the masses, and therefore this ratio would be zero because the $\rho$ contribution would vanish (no isospin violation).

Actually the decays $X\rightarrow J/\psi\rho$ and $X\rightarrow J/\psi\omega$ are not allowed because of phase-space, for $\rho$ and $\omega$ with fixed masses, but can occur when their mass distribution is considered and will be seen in the decays $X\rightarrow J/\psi\pi\pi$ and $X\rightarrow J/\psi\pi\pi\pi$ respectively, where the two and three pion states are the result of the decays of the $\rho$ and $\omega$. Hence to measure the ratio of the $X$ decaying to two and three pions plus a $J/\psi$ one has to multiply the expression in (\ref{ratio}) by the ratio of the phase-space available for the decay of a $\rho$ to two pions divided by the phase-space for the decay of a $\omega$ into three pions:

\begin{widetext}
\be
\frac{{\cal B}(X\rightarrow J/\psi\pi\pi )}{{\cal B}(X\rightarrow J/\psi\pi\pi\pi )}&=&\left(\frac{G_{11}-G_{22}}{{G_{11}+G_{22}}}\right)^2 \frac{\int_0^{\infty } q \mathcal{S}\left(s,m_{\rho },\Gamma _{\rho }\right) \theta \left(m_X-m_{J/\psi }-\sqrt{s}\right) \,
   ds}{\int_0^{\infty } q \mathcal{S}\left(s,m_{\omega },\Gamma _{\omega }\right) \theta \left(m_X-m_{J/\psi }-\sqrt{s}\right) \, ds}
\frac{{\cal B}_\rho}{{\cal B}_\omega} \label{branchtot}
\ee
\end{widetext}
where ${{\cal B}_\rho}$ and ${{\cal B}_\rho}$ are the branching fractions of $\rho$ decaying into two pions ($\sim$ 100 \%) and $\omega$ decaying into three pions ($\sim$ 89 \%), $\theta(y)$ is the Heaviside theta function and $\mathcal{S}\left(s,m,\Gamma\right)$ is the spectral function of the mesons given by:

\be
\mathcal{S}\left(s,m,\Gamma\right)&=&-\frac{1}{\pi} Im\left(\frac{1}{s-m^2+i \Gamma  m}\right)
\ee

From the expression in eq. (\ref{ratio}) one observes that the isospin violation in the decay of the $X$ will be proportional to the square of the difference between the loops with charged and neutral $D$ mesons. Moreover if one looks at figure \ref{fig1} one sees that this difference is maximal at the threshold of the $D^0\bar D^{*0}$, such that the closer the resonance is to that threshold (the smaller the binding energy) the bigger is the isospin violation in the decay of the $X$. If the $X$ is right over the threshold, the value of $R_{\rho/\omega}$, with the loops calculated with dimensional regularization for $\rho$ and $\omega$ fixed masses, is:

\be
R_{\rho/\omega}&=&0.032
\ee

This is a measure of the isospin violation in the decay of the $X$, which is only about 3\% in spite of the fact that we have chosen the conditions to maximize it. This ratio is of the same order of magnitude as the one obtained in \cite{gutsch2} (see eq. (36) of this paper). However, even this small isospin breaking can lead to sizable values of the ratio of eq. (\ref{branchtot}) when one takes into account the mass distributions of the $\rho$ and $\omega$, which provide different effective phase-spaces in this two possible $X$ decays. Thus, using eq. (\ref{branchtot}), which considers explicitly the $\rho$ and $\omega$ mass distributions, we find the branching ratio:

\be
\frac{{\cal B}(X\rightarrow J/\psi\pi^+\pi^-\pi^0 )}{{\cal B}(X\rightarrow J/\psi\pi^+\pi^- )}&=&1.4 \label{res14}
\ee
which is compatible with the value $1.0\pm0.4$ from experiment \cite{bellegj}.


\section{The two C-parity states}

There are six channels with charm and strangeness equal to zero and positive C-parity. We show in table \ref{tab1} the couplings of the pole obtained solving the scattering equation for these channels.

\begin{table}
\begin{center}
\caption{Couplings of the pole at (3871.6-$i$0.001) MeV to the channels ($\alpha_H$=-1.27 here).} \label{tab1}
\begin{tabular}{c|c}
\hline
Channel & $|g_{R\rightarrow PV}|$ [MeV] \\
\hline
\hline
 & \\
$\pi^-\rho^+-c.c.$ & 1.4 \\
\hline
 & \\
$K^- \bar K^{*+}-c.c.$ & 8.7 \\
\hline
 & \\
$K^0 \bar K^{*0}-c.c.$ & 7.4 \\
\hline
 & \\
$D^- \bar D^{*+}-c.c.$ & 2982 \\
\hline
 & \\
$D^0 \bar D^{*0}-c.c.$ & 3005 \\
\hline
 & \\
$D_s^- \bar D_s^{*+}-c.c.$ & 2818 \\
\hline
\end{tabular}
\end{center}
\end{table}

One can see in table 1 that, although there is some isospin violation in the couplings, it is very small, less than 1 \%. One might think that if the binding energy is much smaller than the difference between the neutral and charged thresholds (8 MeV), the resonance will be mostly dominated by the neutral channel, the one closest to the threshold. The binding energy in the case of the pole in Table \ref{tab1} is 0.4 MeV. As we mentioned, in the limit that the binding energy goes to zero, the couplings should all vanish. We show in figure \ref{plotcoup1} that, indeed, the coupling of the $X$ to $D^0\bar D^{*0}$goes to zero for small binding energies, and in figure \ref{plotdiff} we show that even though the difference between the neutral and charged couplings grows for small binding energies, they are of the same order of magnitude. The wave function of the $X(3872)$ is, thus, very close to the isospin I=0 combination of $D^0\bar D^{*0}-c.c.$ and $D^-\bar D^{*+}-c.c.$ and has a sizable fraction of the $D_s^-D_s^{*+}-c.c.$ state.

\begin{figure}[h]
\begin{center}
\begin{tabular}{c}
\includegraphics[width=5cm,angle=-90]{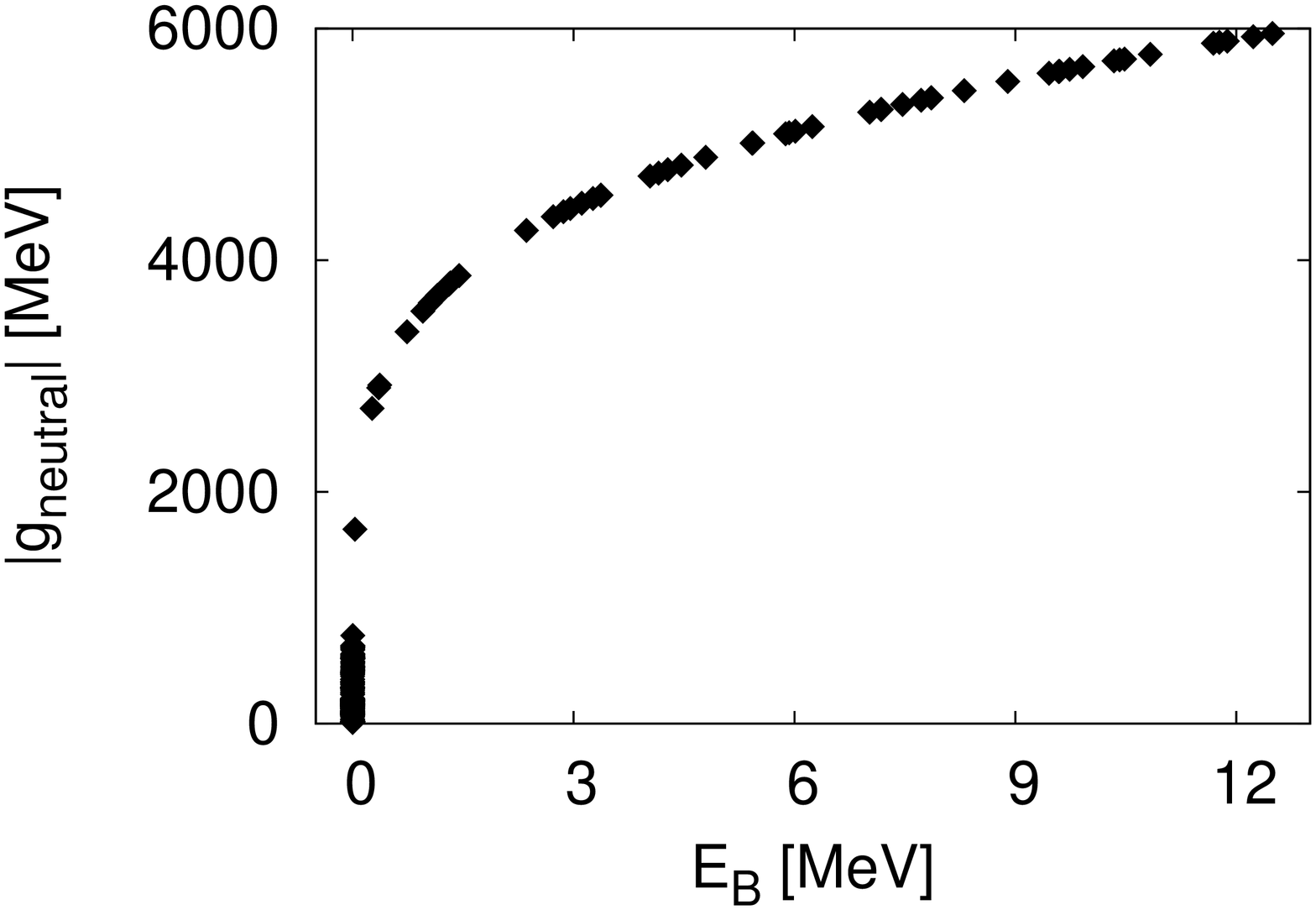} \\
\includegraphics[width=5cm,angle=-90]{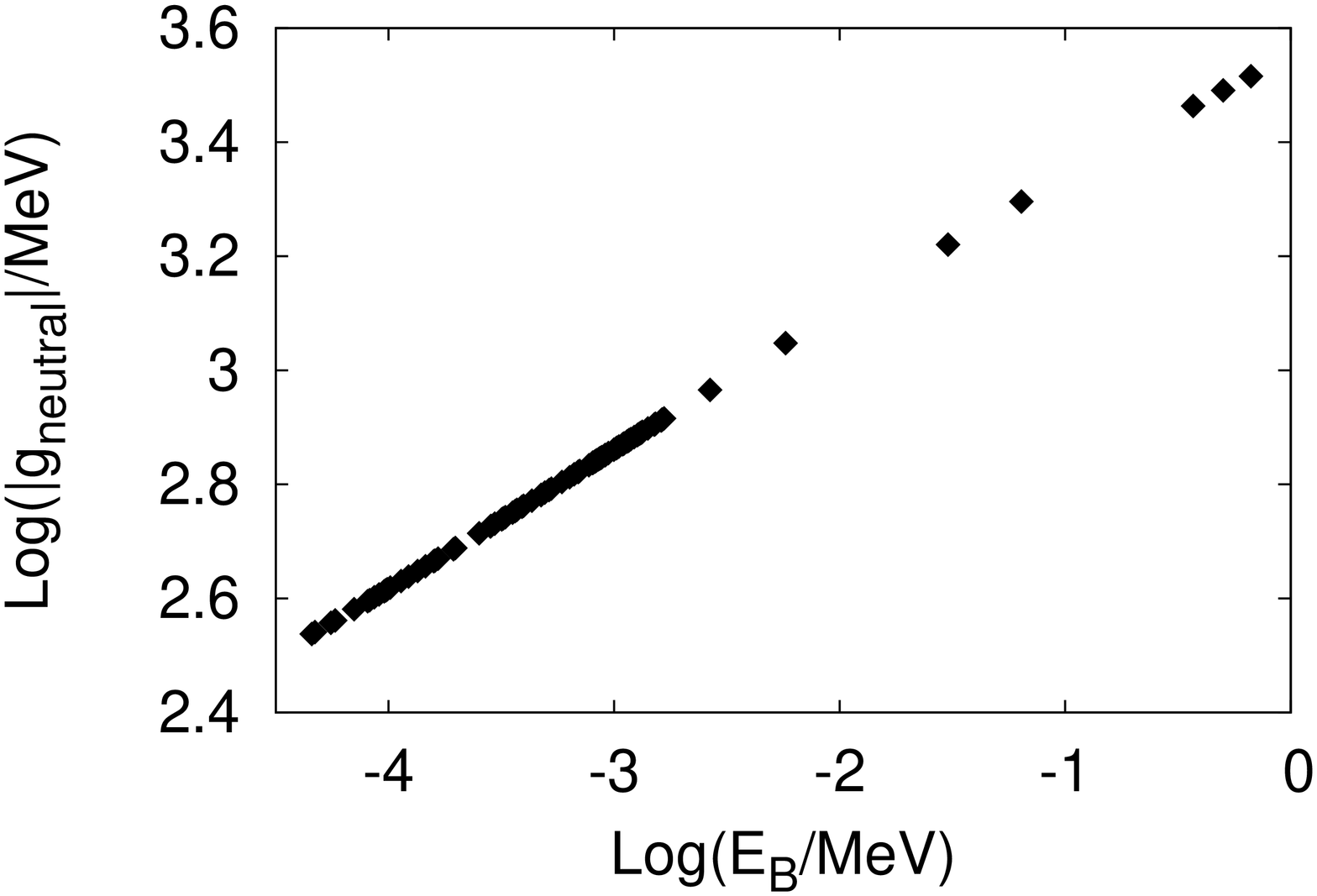} 
\end{tabular}
\caption{Coupling of the $X$ to the $D^0\bar D^{*0}$ channel for different biding energies.} \label{plotcoup1}
\end{center}
\end{figure}

\begin{figure}[h]
\begin{center}
\includegraphics[width=5cm,angle=-90]{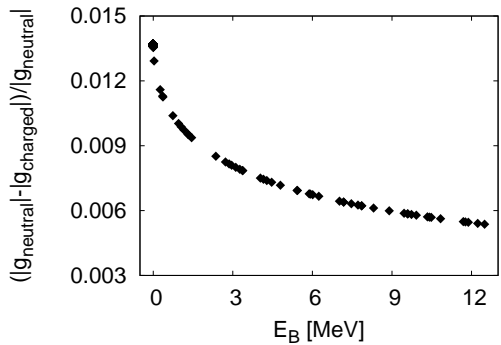} 
\caption{Difference of the coupling of the $X$ with neutral and charged $D\bar D^*$ channels.} \label{plotdiff}
\end{center}
\end{figure}

From figure \ref{plotdiff} we notice that indeed the isospin violation in the couplings of the $X$ to the $D\bar D^*$ channels is bigger for small binding energies, but it reaches a maximum of about 1.4\% which is a very small value. We can go back to the argument that lead to eq. (\ref{ratio}) and the only difference would be that the $G_{11}$ and $G_{22}$ functions would be multiplied by the $D^0\bar D^{*0}-c.c.$ and $D^+\bar D^{*-}$ couplings from table \ref{tab1}, which barely affect the results obtained in eq. (\ref{res14}), since the differences in the couplings are much smaller than those between $G_{11}$ and $G_{22}$. Although a $D^0\bar D^{*0}+c.c.$ state is proposed for the $X(3872)$ in \cite{braaten2}, a formalism accounting for the charged component of this resonance is also presented in \cite{braaten3}. Our results would correspond to taking a value of the parameter $\gamma_1$ much bigger than $\kappa_1(0)$ in size in eq. (39b) of \cite{braaten3}. However, no claims for any particular value of $\gamma_1$ are made in \cite{braaten3}, where only the formalism is presented.

As we already mentioned, we find a second state with negative C-parity. Some of the 22 channels with negative C-parity
have isospin I=0 to which the resonance can decay. There are also pure isospin I=1 channels but, although the generated resonance is an isospin I=0 state, these isospin I=1 channels will couple to it since we are considering here some amount of isospin violation coming from the different masses of charged and neutral members of a same isospin multiplet.

For values of $\alpha_H$ similar to those used in the generation of the $X(3872)$ ($\alpha_H$-1.27), the pole with negative C-parity is in the wrong Riemann sheet, but its effects can still be seen in the cross sections of some channels. We show in figure \ref{figcross} the $|T|^2$ plots of some channels. By taking smaller values of $\alpha_H$ (around -1.36) one can recover a pole below threshold with $\sqrt s$=(3871.4-i26.2) MeV. In our previous work of \cite{danielaxial} this state was narrower. The reason for its relative big width in the present work is the inclusion of the $\eta$-$\eta\prime$ mixing. As was explained in \cite{withzou} the hidden charm dynamically generated states that we obtain are $SU(3)$ singlets and if one considers only the mathematical $\eta_8$ without its mixing with a $SU(3)$ singlet state $\eta_1$, the open channels for the resonance to decay are $SU(3)$ octets and are therefore suppressed. Only when considering also this singlet and hence the physical $\eta$ and $\eta\prime$ states the open channels acquire a $SU(3)$ singlet component to which the resonance strongly couples.

\begin{figure}[h]
\begin{center}
\begin{tabular}{cc}
\includegraphics[width=2.5cm,angle=-90]{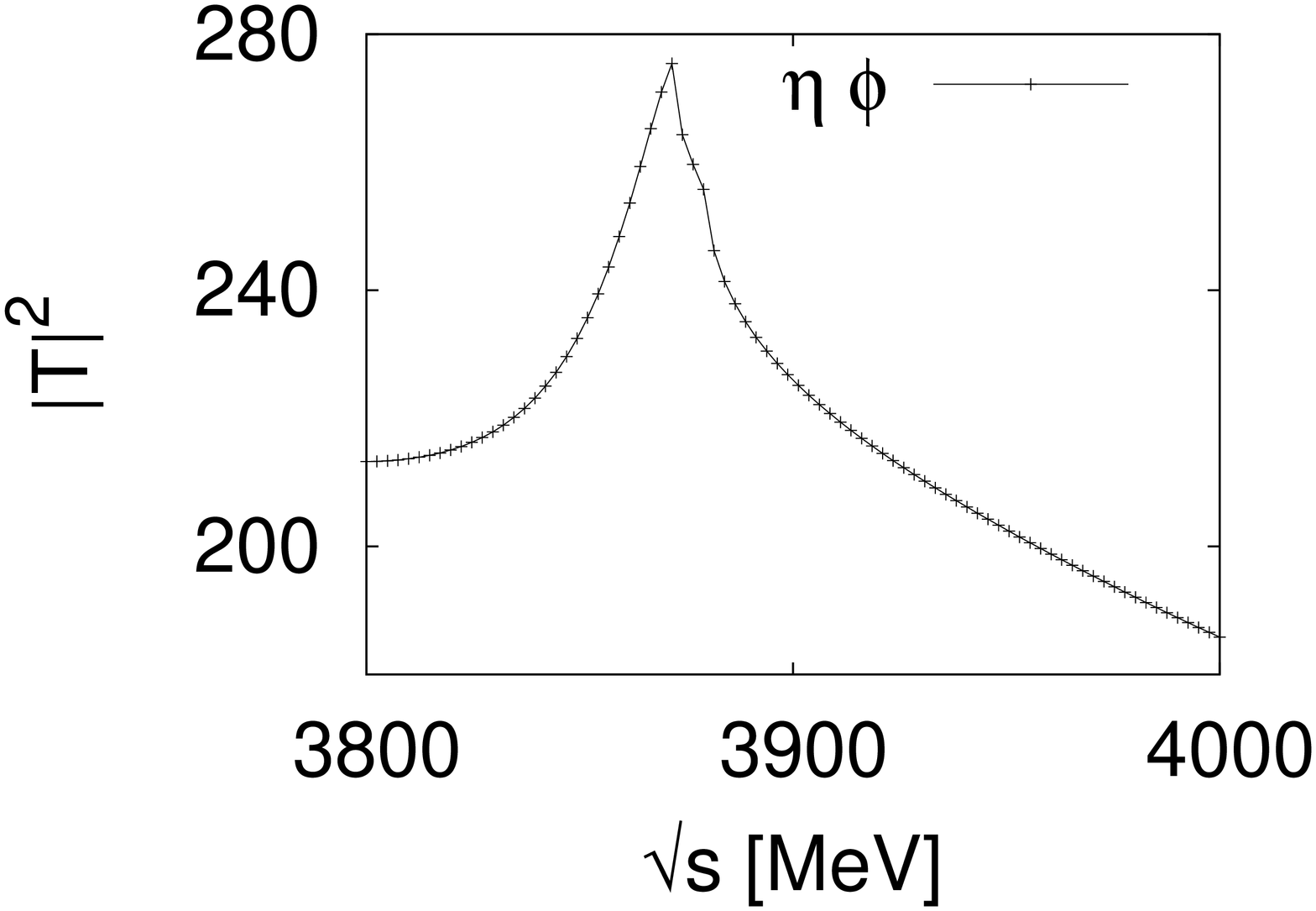} & \includegraphics[width=2.5cm,angle=-90]{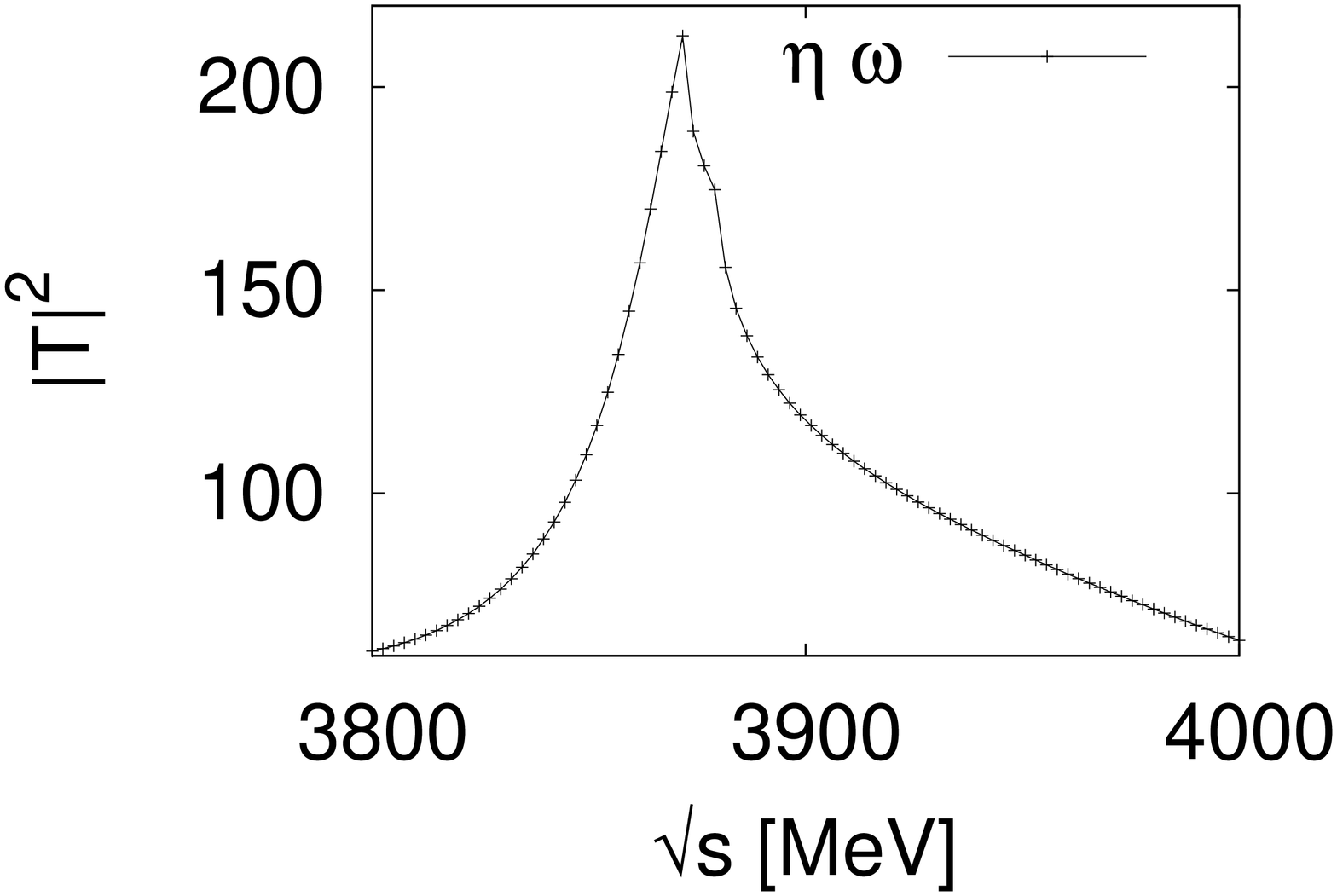} \\
\includegraphics[width=2.5cm,angle=-90]{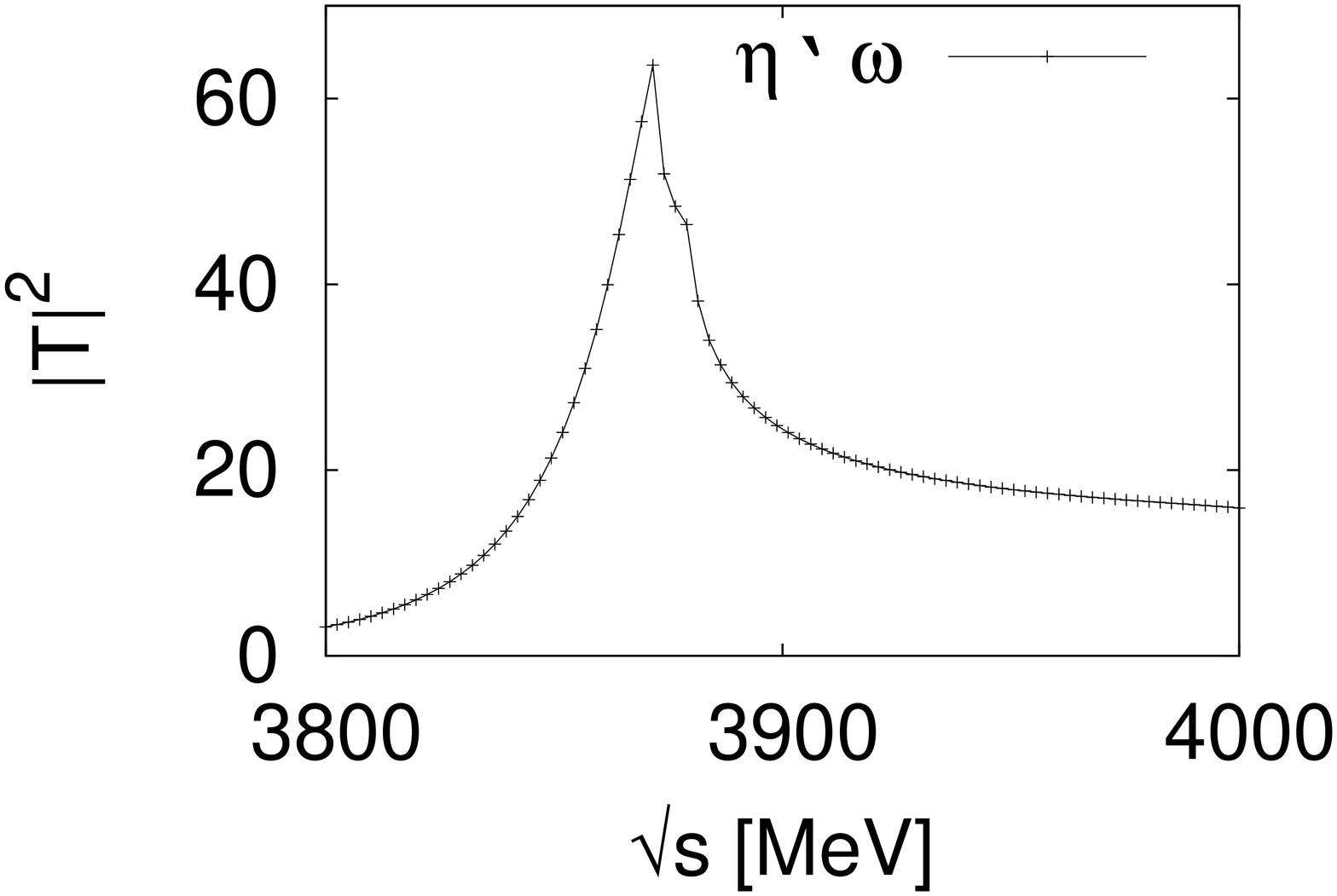} & \includegraphics[width=2.5cm,angle=-90]{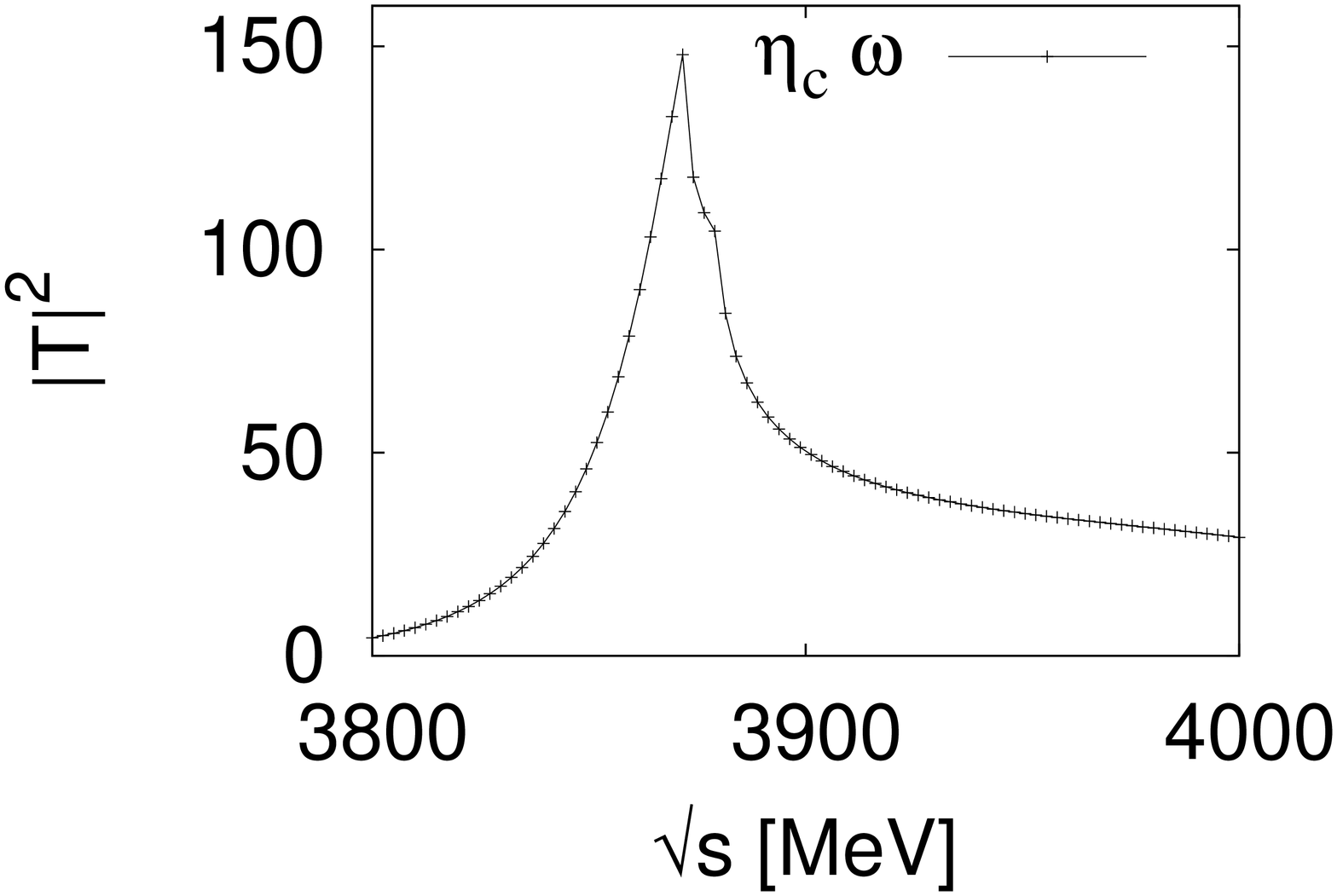} 
\end{tabular}
\caption{The $|T|^2$ plot for some of the negative C-parity channels.} \label{figcross}
\end{center}
\end{figure}

The channels shown in figure \ref{figcross} are those where there is phase-space available for the resonance to decay and to which it couples most strongly.

Next we want to compare the results obtained from our approach with the experiment data from \cite{expcross} for $B\rightarrow KD^0\bar D^{*0}$. For this we follow the approach of \cite{hanhart,danielaxial} where one shows that the experimental data for $d\Gamma/dM_{inv}(D \bar D^{*})$ are proportional to $p|T_{D\bar D^*\rightarrow D\bar D^*}|^2$ where $p$ is the center of mass momentum of the $D$ meson with $M_{inv}$ invariant mass.

We show in figure \ref{crossexp} plots of the $d\Gamma/dM_{inv}$ for the channels $D^0\bar D^{*0}\pm c.c.$ and the pure $D^0\bar D^{*0}$ and compare it with experimental data from \cite{expcross}.

\begin{figure}[h]
\begin{center}
\begin{tabular}{c}
\includegraphics[width=5cm,angle=-90]{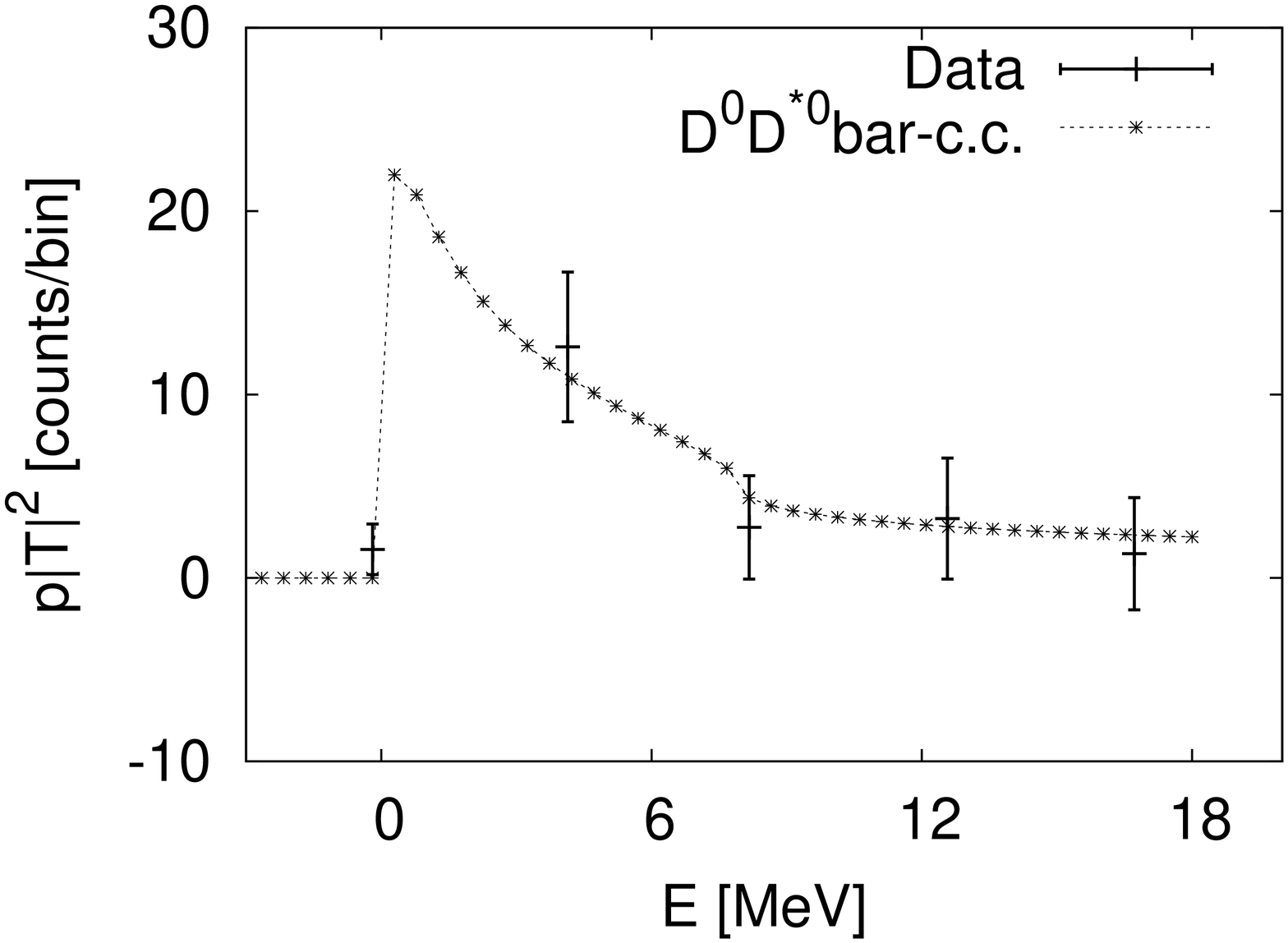} \\
\includegraphics[width=5cm,angle=-90]{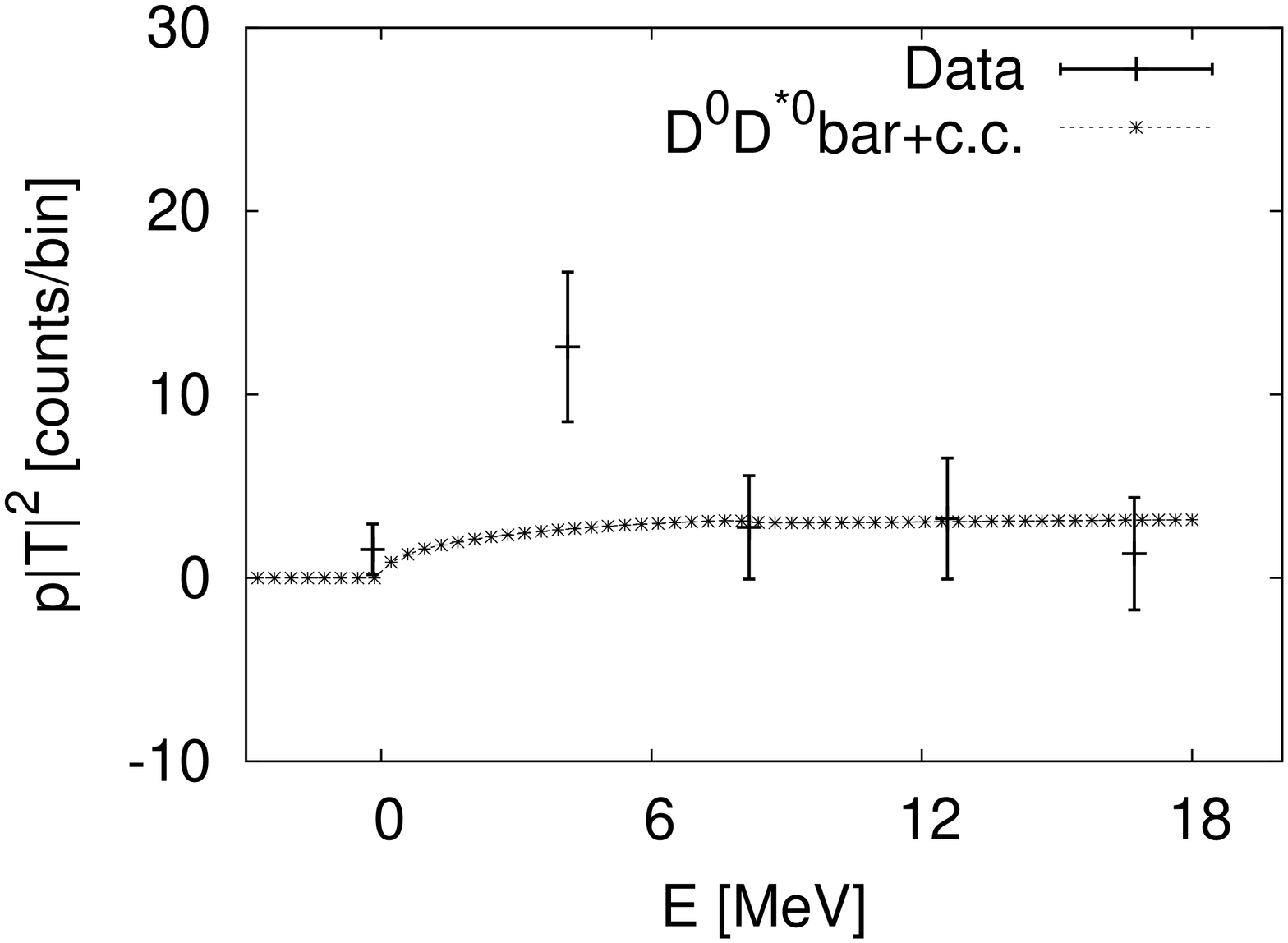} \\
\includegraphics[width=5cm,angle=-90]{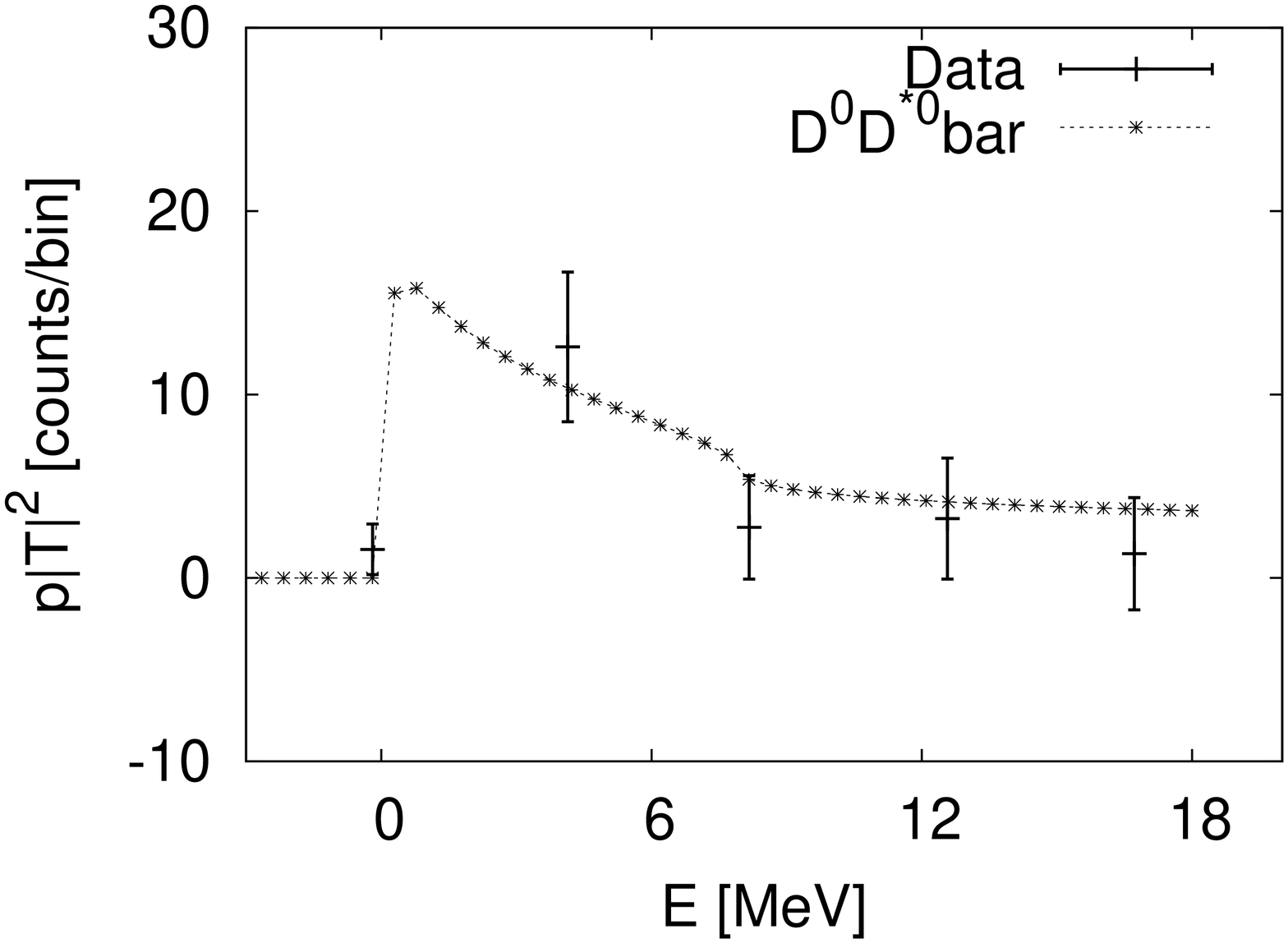} 
\end{tabular}
\caption{The $p|T|^2$, where $p$ is the three momentum of the $D$ mesons and $E$ is the energy above threshold, compared to data from \cite{expcross}. $\alpha=-$ 1.27 here.} \label{crossexp}
\end{center}
\end{figure}

In the plots of figure \ref{crossexp} the theoretical curves have been normalized to fit the experimental data. One can clearly see that the positive C-parity state alone describes the data while the negative C-parity state alone does not describes it. However this experiment can not determine whether the $D^0\bar D^{*0}$ (together with the $\bar D^0D^{*0}$ summed incoherently) comes from a given C-parity. In the lower plot of figure \ref{crossexp} we evaluate the differential cross section for the state $D^0\bar D^{*0}$, which has contribution from both C-parity states. As we can see in the figure, the results obtained are also in agreement with the data and, hence, in spite of the results in the middle plot of figure \ref{crossexp}, the experimental data does not rule out the existence of the negative C-parity state.

The former discussions put the two states that we predict in a perspective concerning the $D^0\bar D^{*0}$ production experiment. In what follows we are going to do a more subtle exercise to bring some light into a current discussion on whether the combination of the data on $X\rightarrow J/\psi \pi \pi$ and $X\rightarrow D^0\bar D^{*0}$ reactions determine if the state $X(3872)$ is a bound state or a virtual one. In what follows we are going to consider only the contribution from the positive C-parity state. In \cite{braaten2} a slightly bound state is preferred, although a virtual state is not ruled out, while in \cite{hanhart} a virtual state is claimed. With our detailed description of coupled channels, our approach is in a favorable position to get into the debate and bring new information. Yet, to do so one needs to introduce two new elements into consideration: the width of the $D^{*0}$ meson and the smearing of the results with experimental resolution. This was claimed to be relevant in \cite{voloshin2} and \cite{braaten2}. We have considered this by taking for the $D^{*0}$ width $\Gamma_{D^{*0}}$=65 KeV as in \cite{braaten2} and the experimental resolution $\Delta E$=2.5 MeV. The consideration of the width of the $D^{*0}$ is taken into account by folding the $D^0\bar D^{*0}$ loop function $G$ with the spectral function of the $D^*$ meson,

\be
{\cal S}(\tilde{M})&=&\left({-1\over\pi}\right) Im{1\over \tilde{M}^2-M_{D^*}^2+iM_{D^*}\Gamma_{D^*}}, \label{spectral}
\ee
in the calculation of the T-matrix, as done in eq. (20) of \cite{danielaxial}. On the other hand the result for $\frac{d\Gamma}{dM_{inv}}$ is folded with the mass distribution of the $D^*$ of eq. (\ref{spectral}), since in the phase-space the three momentum $q$ of the $D^0\bar D^*$ system appears as a factor and this three momentum depends on the mass of the $D^*$. The final result is folded by a Gaussian distribution with a width of 2.5 MeV to simulate the experimental resolution. In this way one gets strength below the nominal threshold of $D^0\bar D^{*0}$ for the decay of the $X(3872)$ into $D^0\bar D^{*0}$.

With these considerations we change slightly the $\alpha$ parameter which governs whether we obtain a bound state or a slightly unbound, virtual state. We normalize the two invariant mass distributions to the experimental data. The shapes alone tell us which option is preferable.

In figures \ref{figcomp1} and \ref{figcomp2} we show the results for the different values of $\alpha$. To the left we have the results for $X\rightarrow J/\psi \pi \pi$ and to the right those for $X\rightarrow D^0\bar D^{*0}$. What we see is that the effect of the convolution with the $D^*$ width and the experimental resolution is important, as claimed in \cite{voloshin2, braaten2} and help us make a choice of the preferred situation. At simple eye view, corroborated by a $\chi^2$ evaluation, see table \ref{tabchi2}, the preferred combined solution corresponds to $\alpha=-$1.23 for which we have a slightly unbound, virtual state. This is the preferred solution in \cite{hanhart}, also not ruled out in \cite{braaten2}.

\begin{widetext}
\begin{figure}
\begin{tabular}{cc}
\includegraphics[width=6cm,angle=-90]{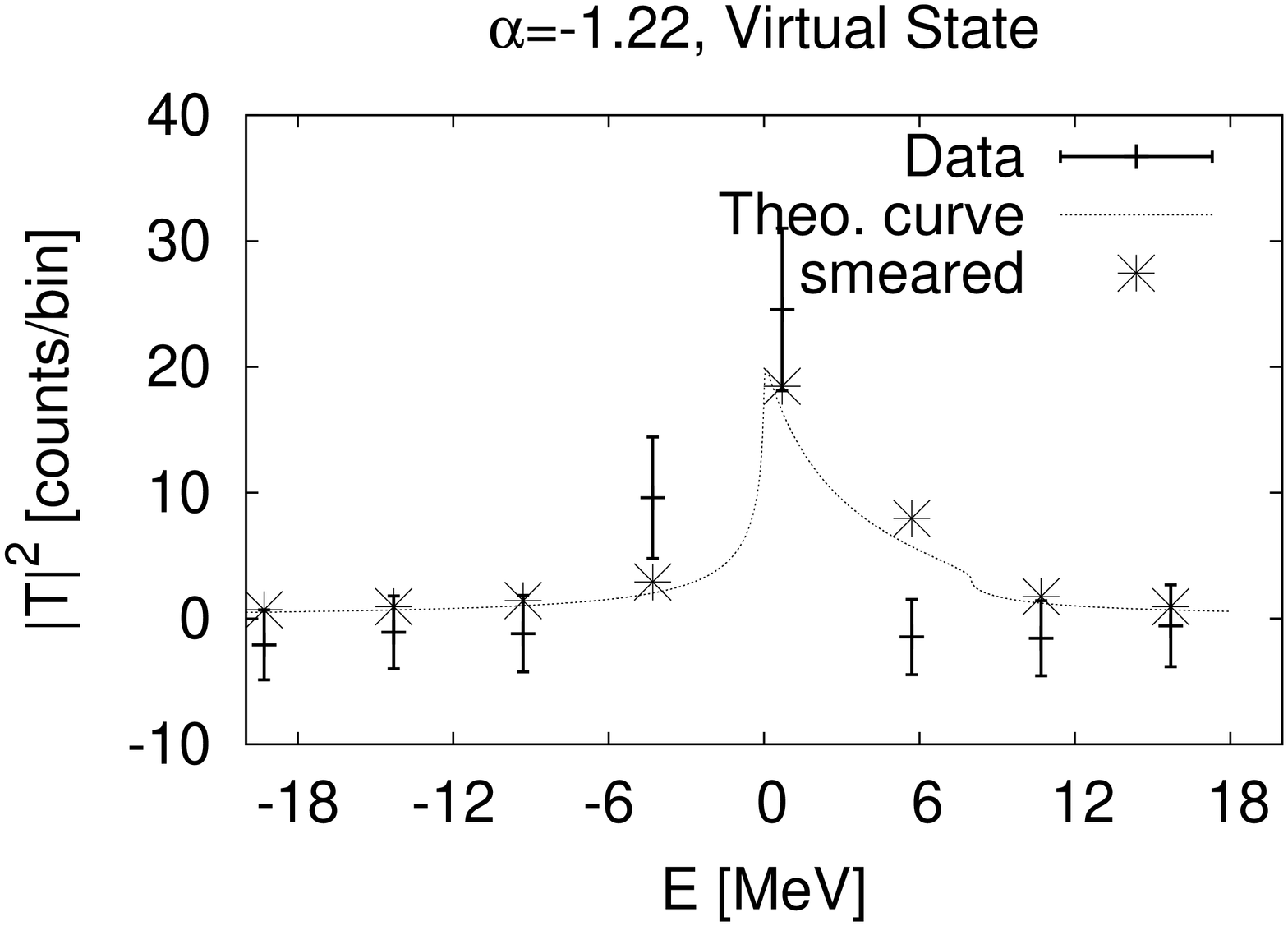} & \includegraphics[width=6cm,angle=-90]{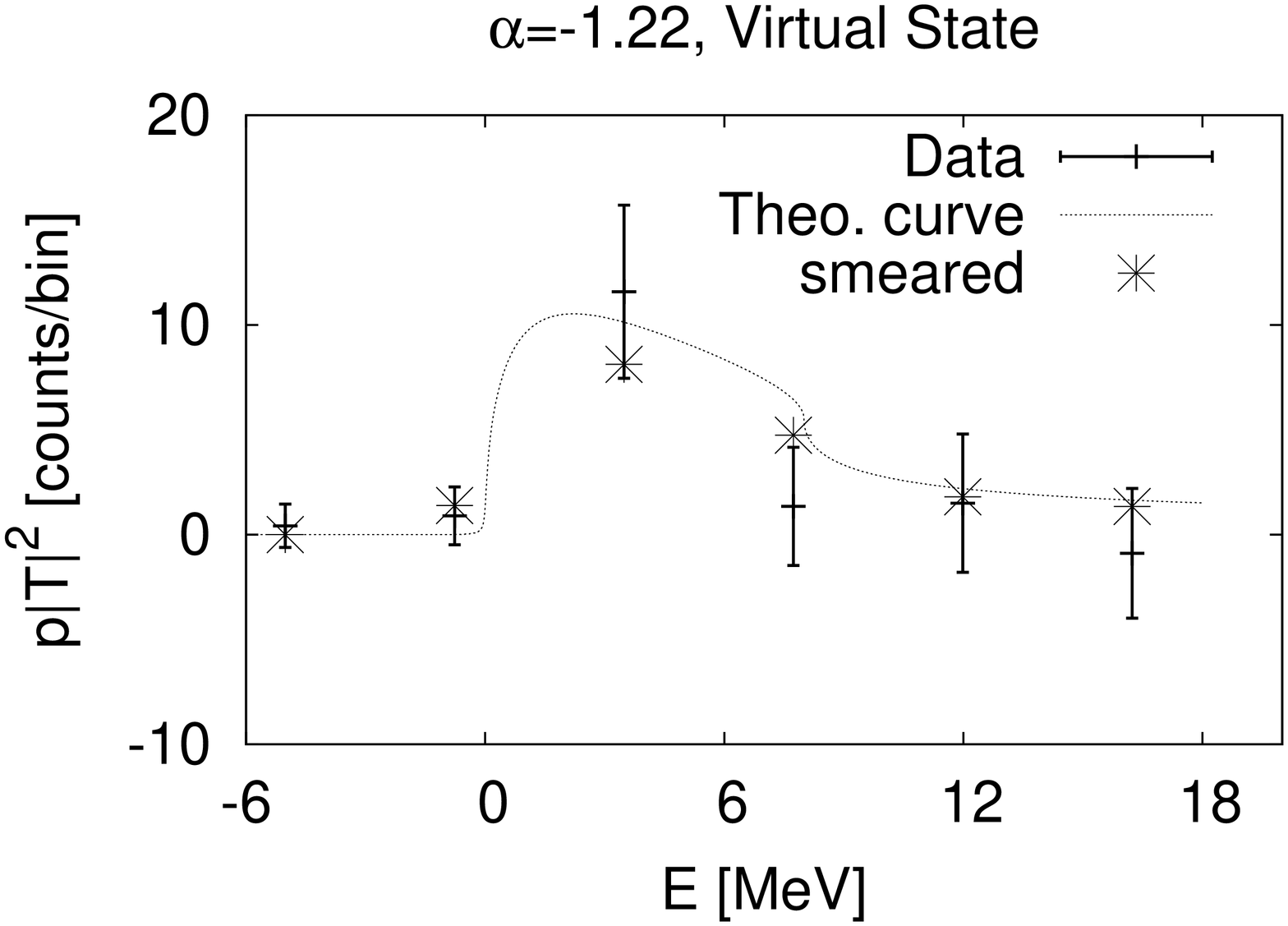} \\
\includegraphics[width=6cm,angle=-90]{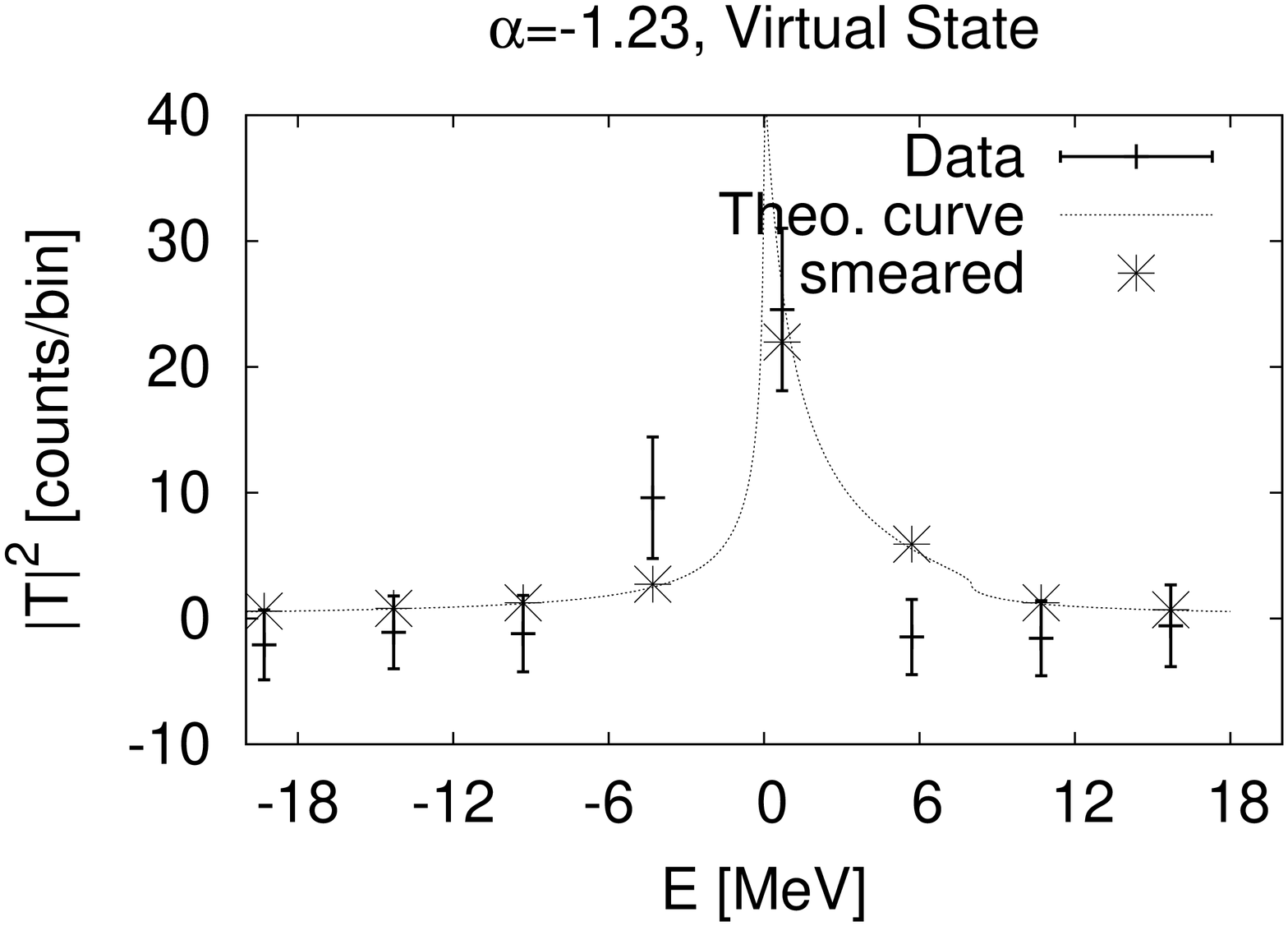} & \includegraphics[width=6cm,angle=-90]{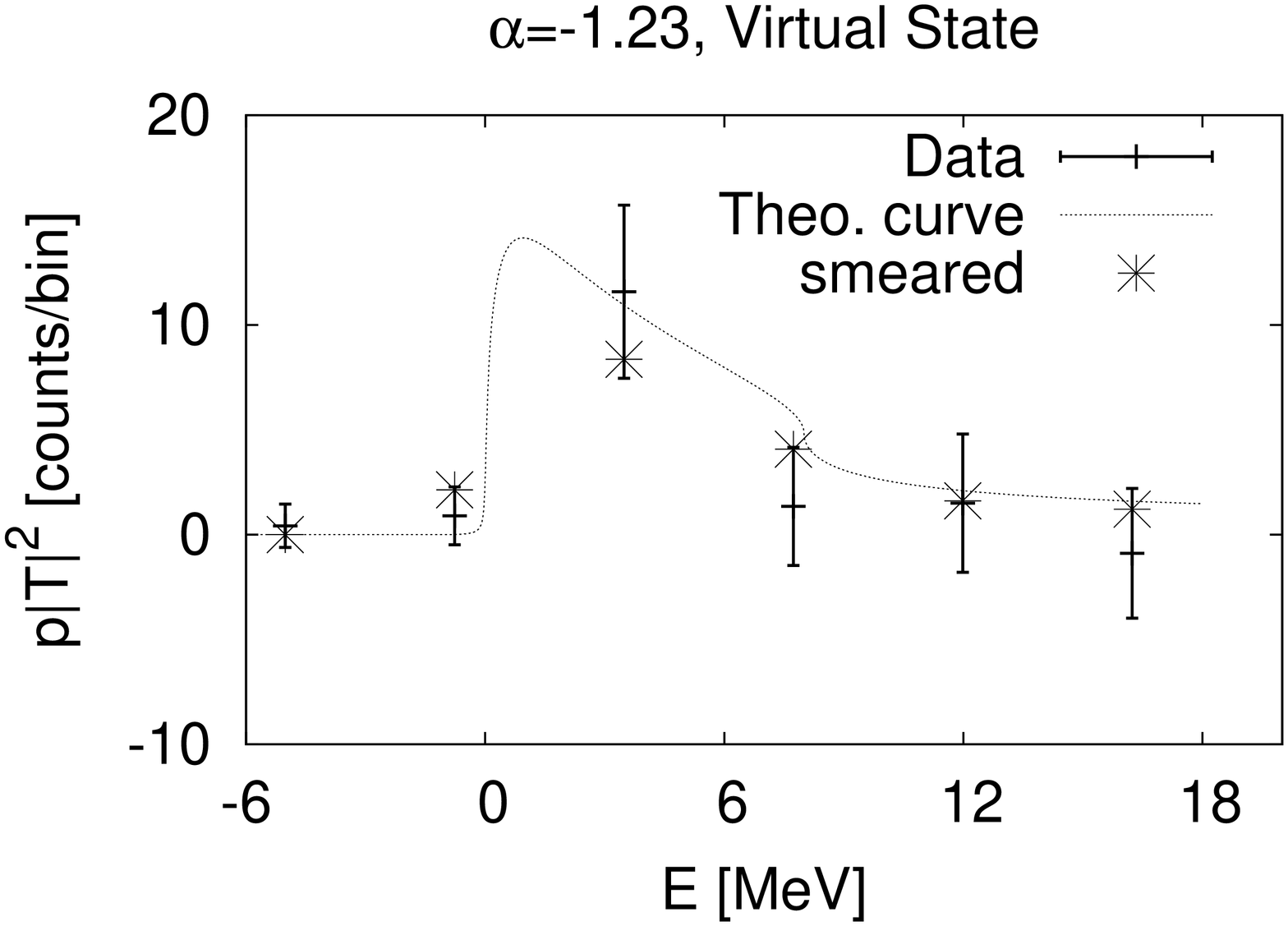} \\
\includegraphics[width=6cm,angle=-90]{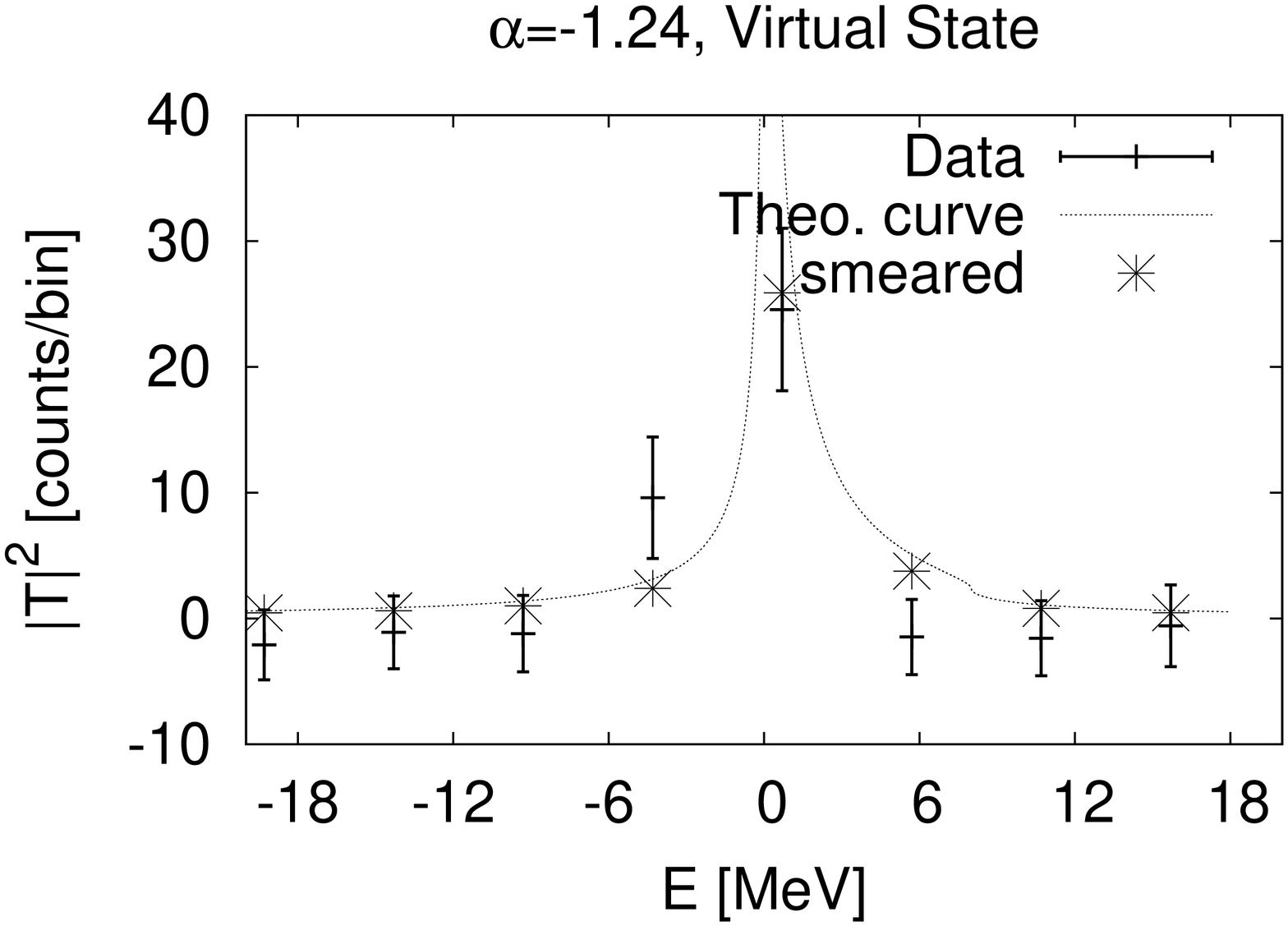} & \includegraphics[width=6cm,angle=-90]{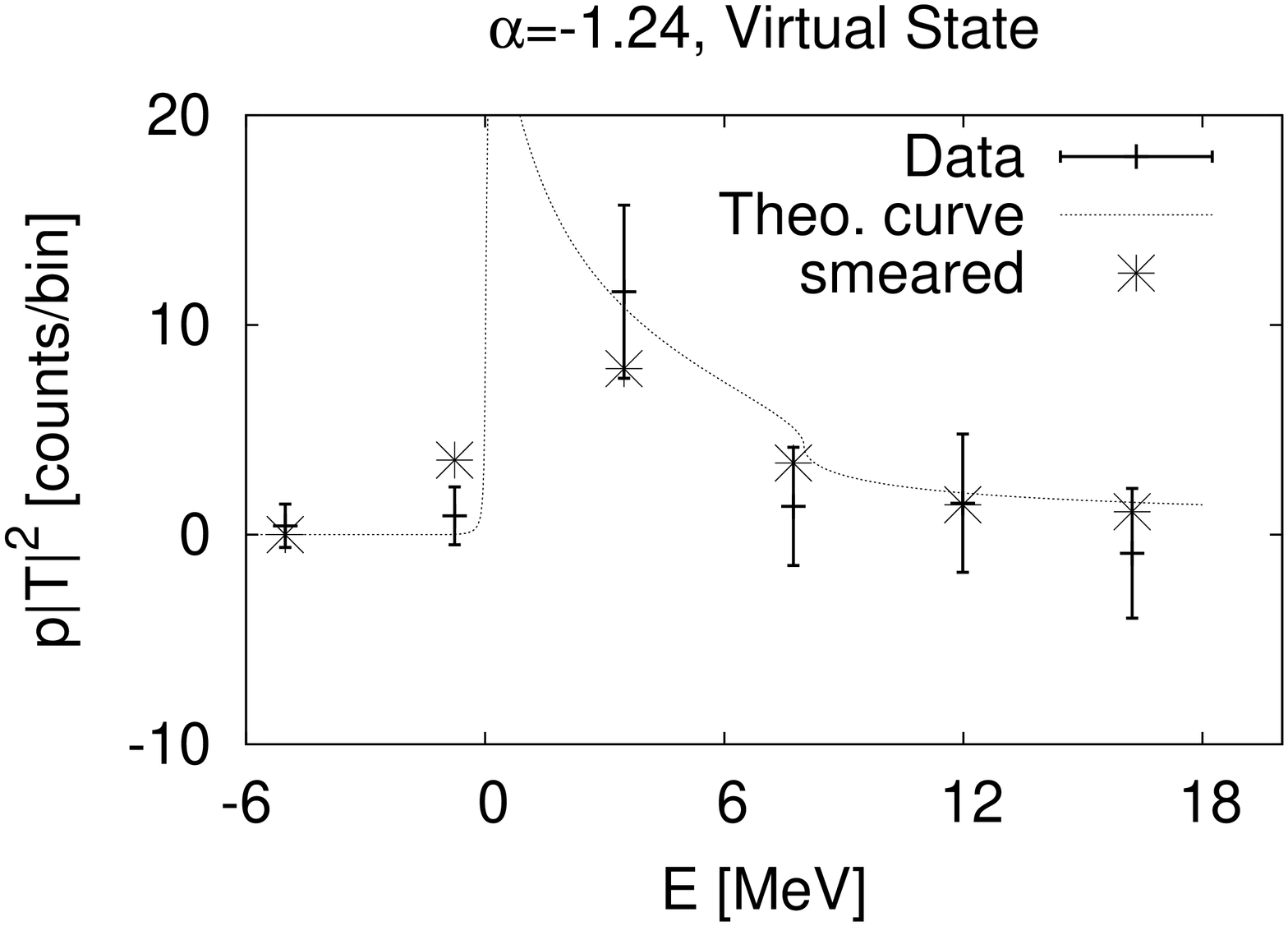} \\
\end{tabular}
\caption{Theoretical results compared to data from \cite{expcross}. The smeared points are calculated from the theoretical curve by folding it with a Gaussian, simulating the experimental resolution.} \label{figcomp1}
\end{figure}
\end{widetext}

\begin{widetext}
\begin{figure}
\begin{tabular}{cc}
\includegraphics[width=6cm,angle=-90]{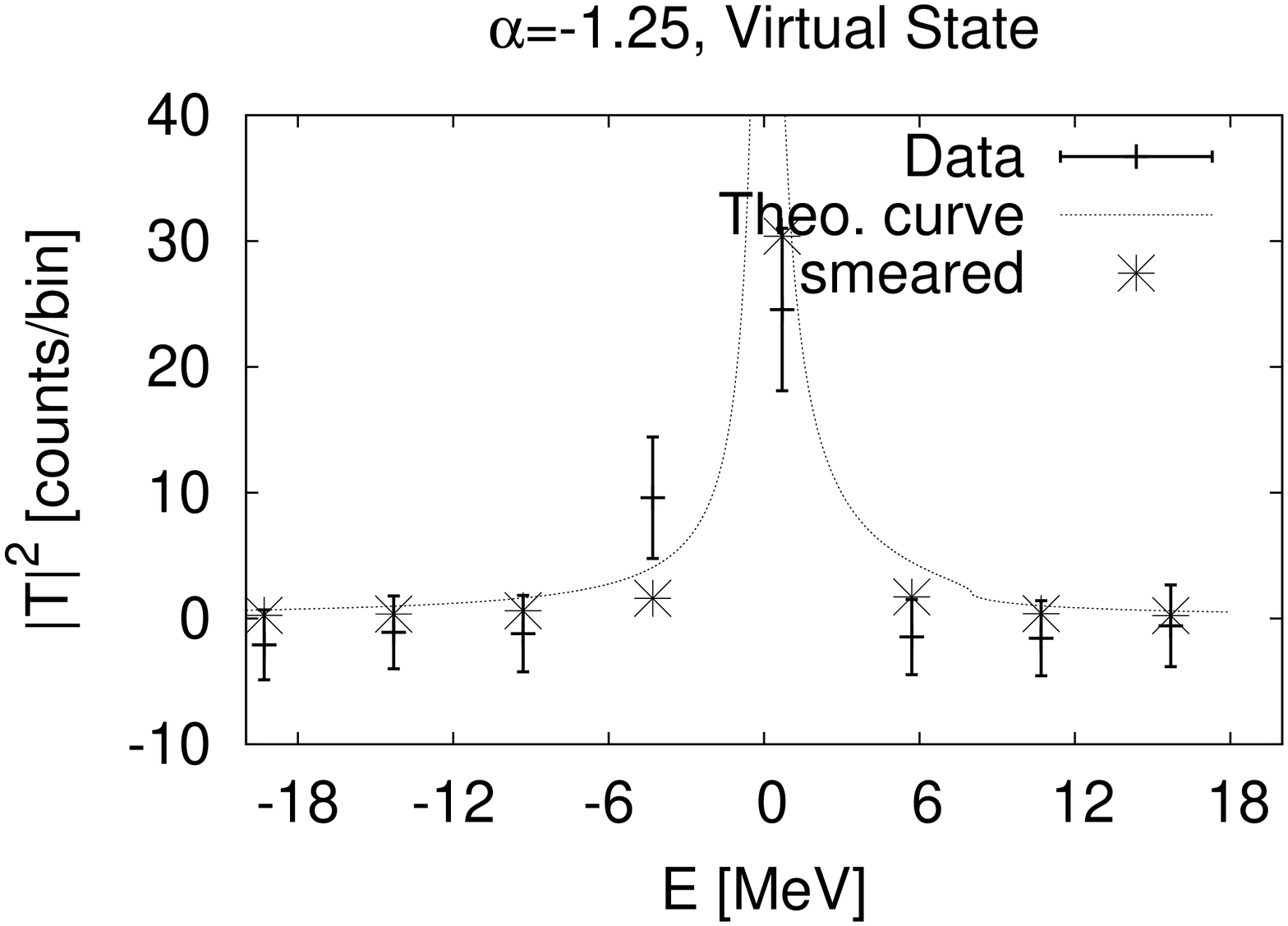} & \includegraphics[width=6cm,angle=-90]{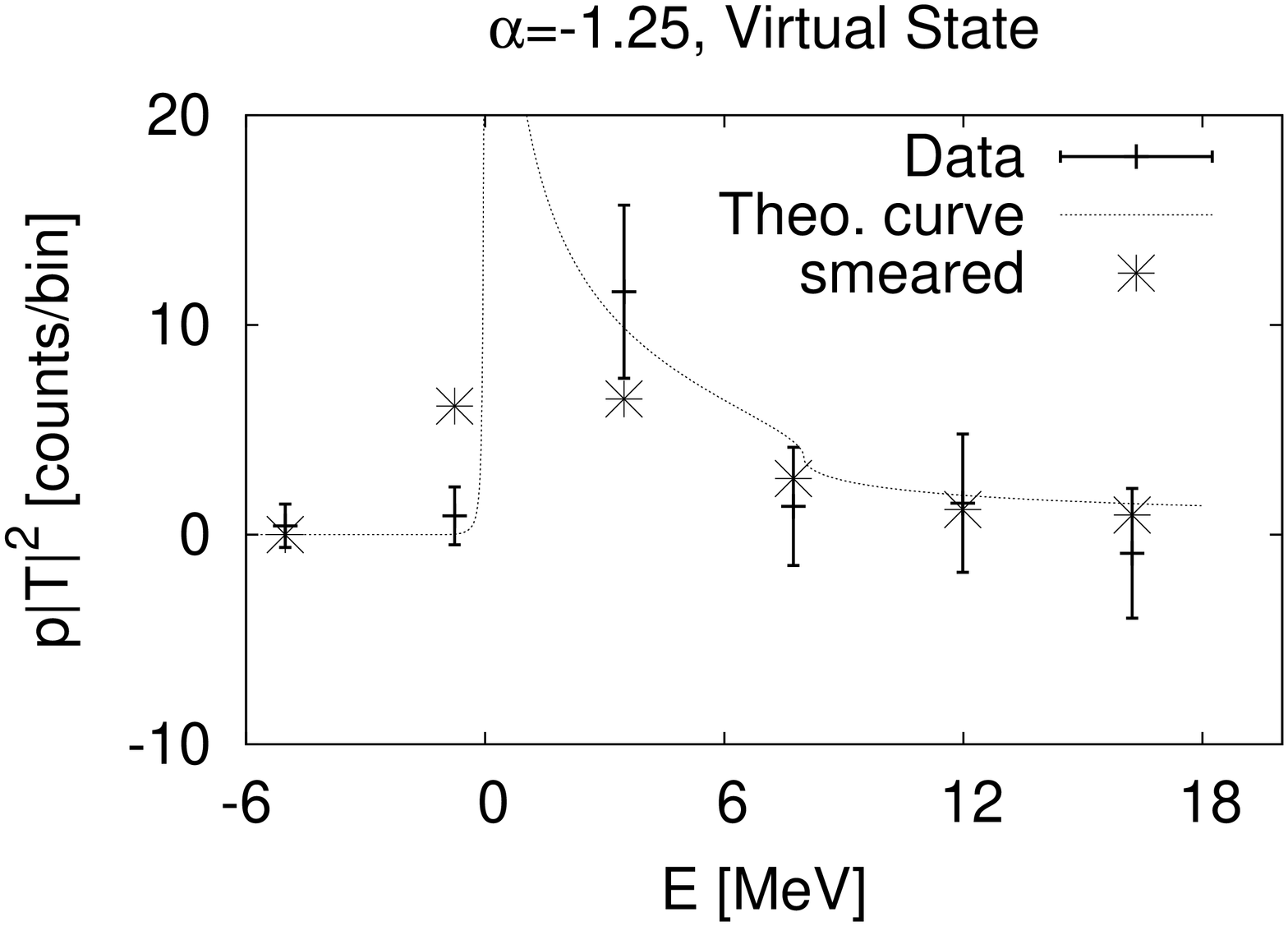} \\
\includegraphics[width=6cm,angle=-90]{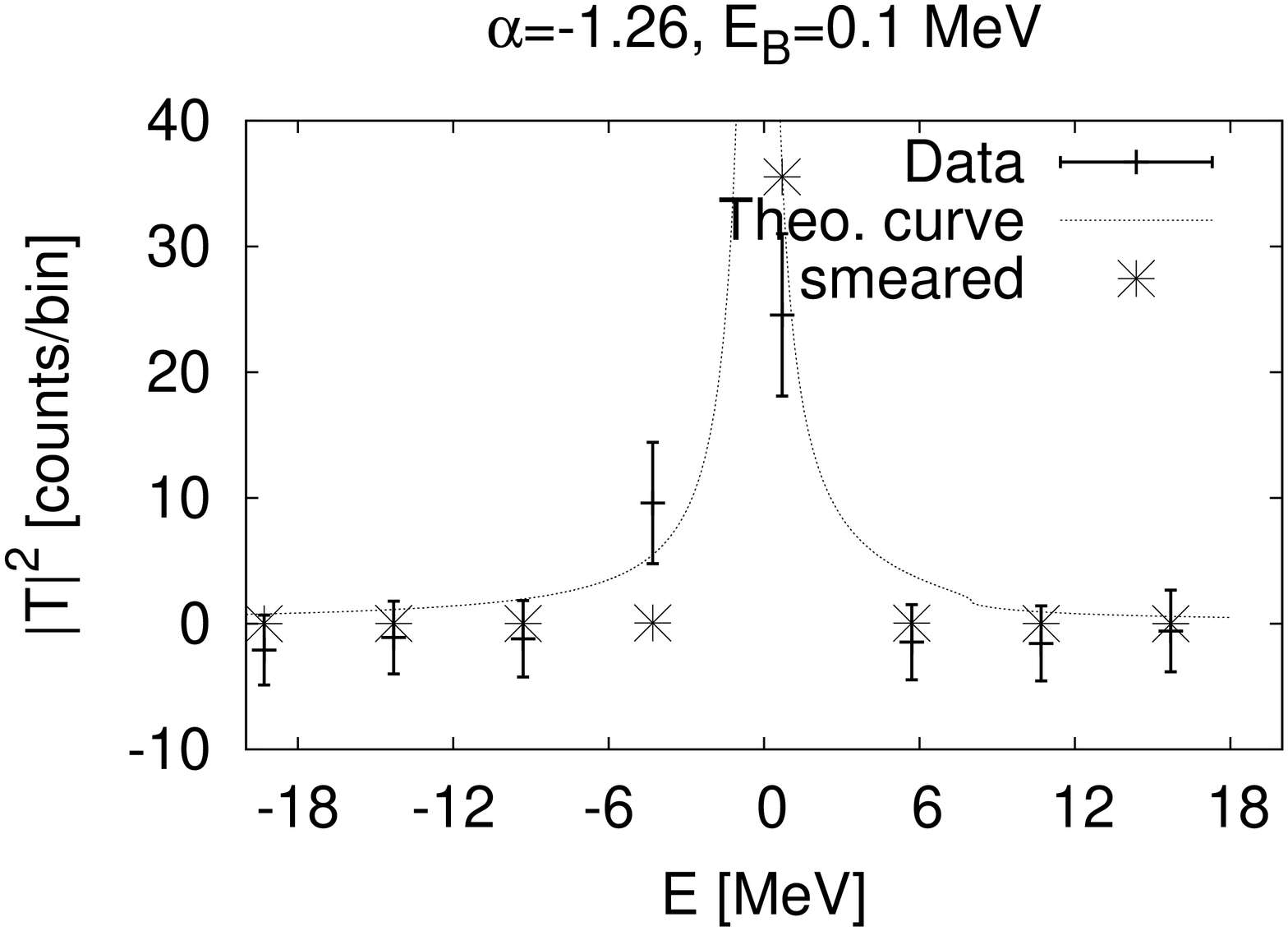} & \includegraphics[width=6cm,angle=-90]{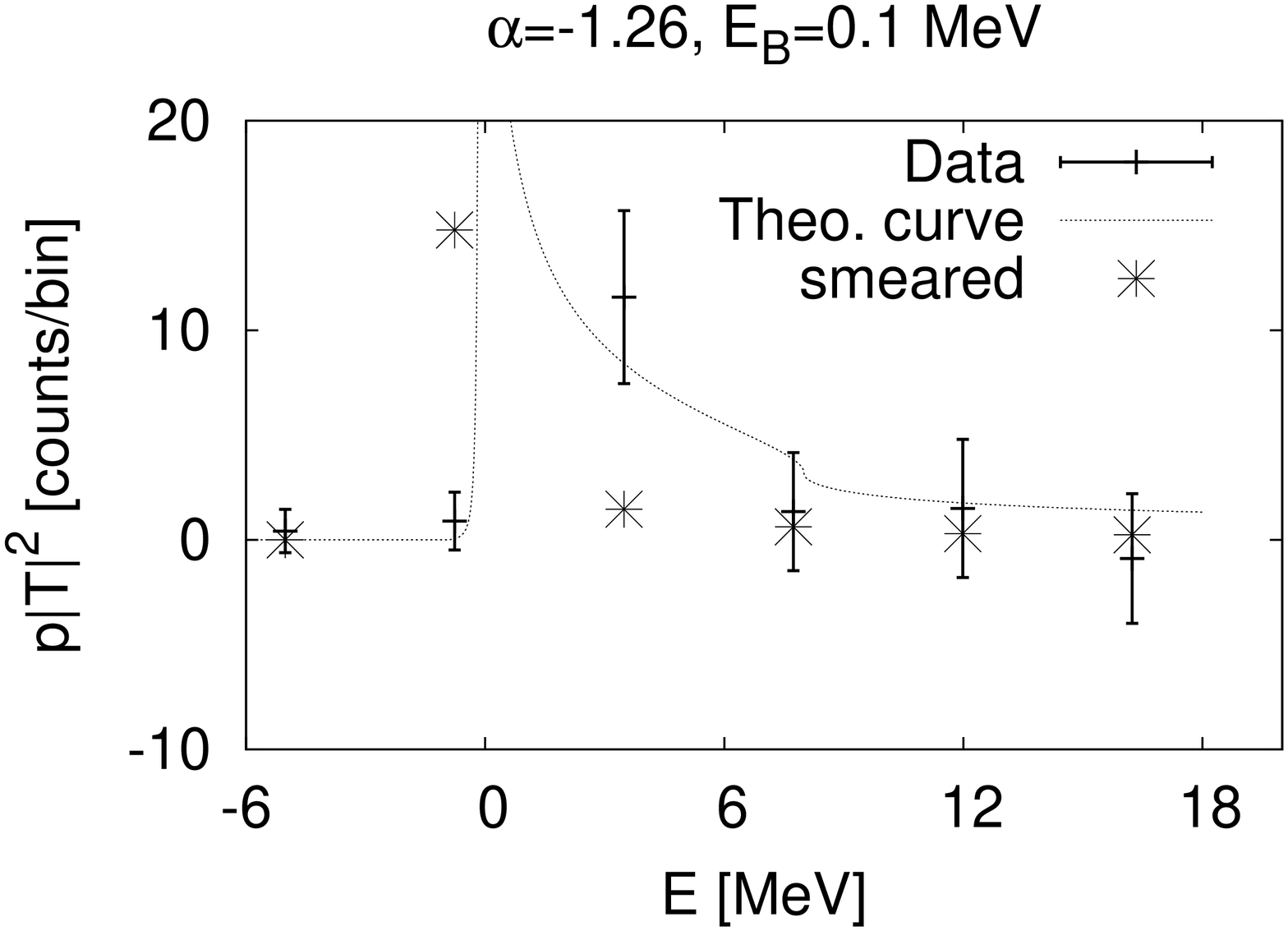} \\
\includegraphics[width=6cm,angle=-90]{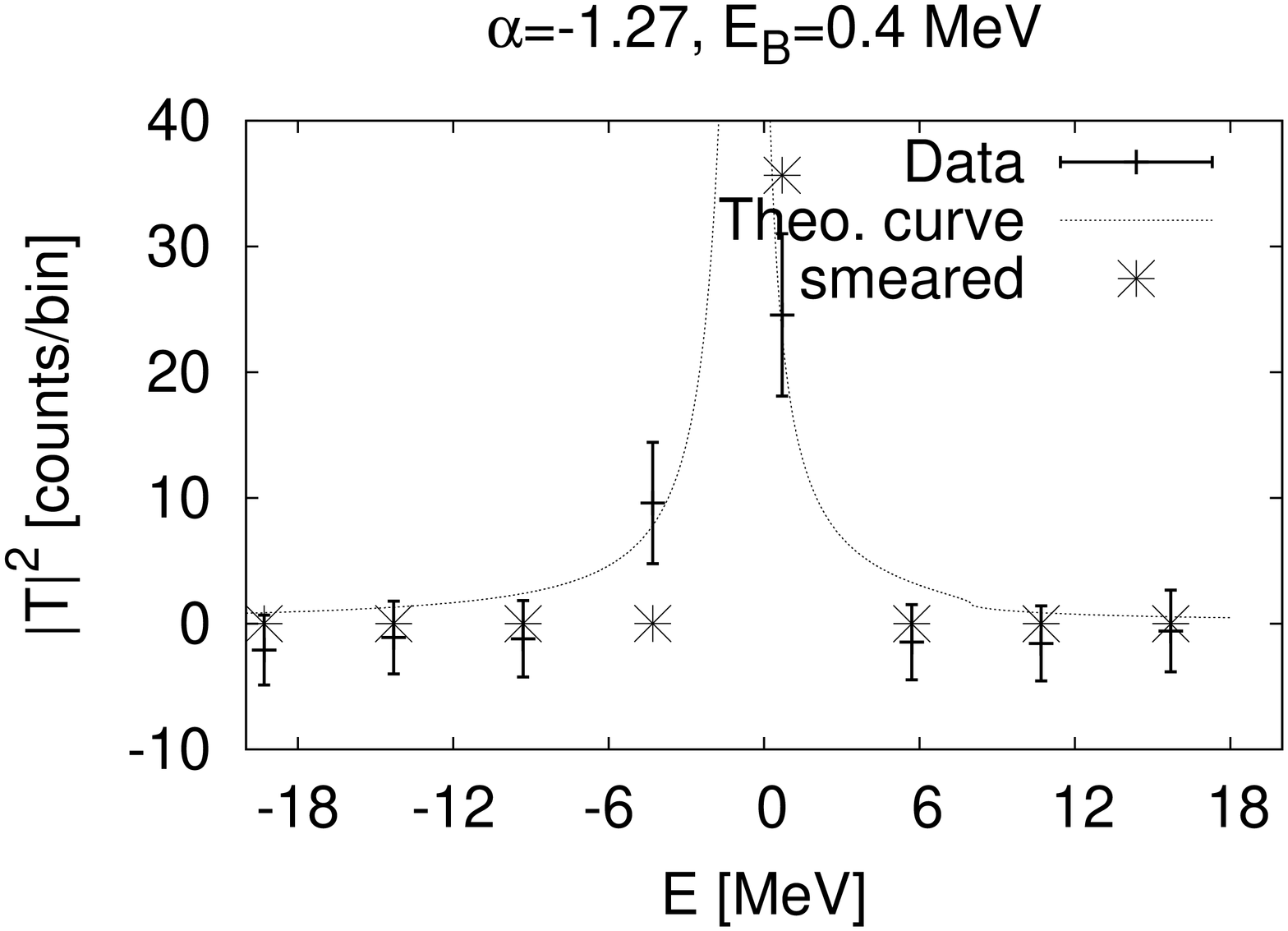} & \includegraphics[width=6cm,angle=-90]{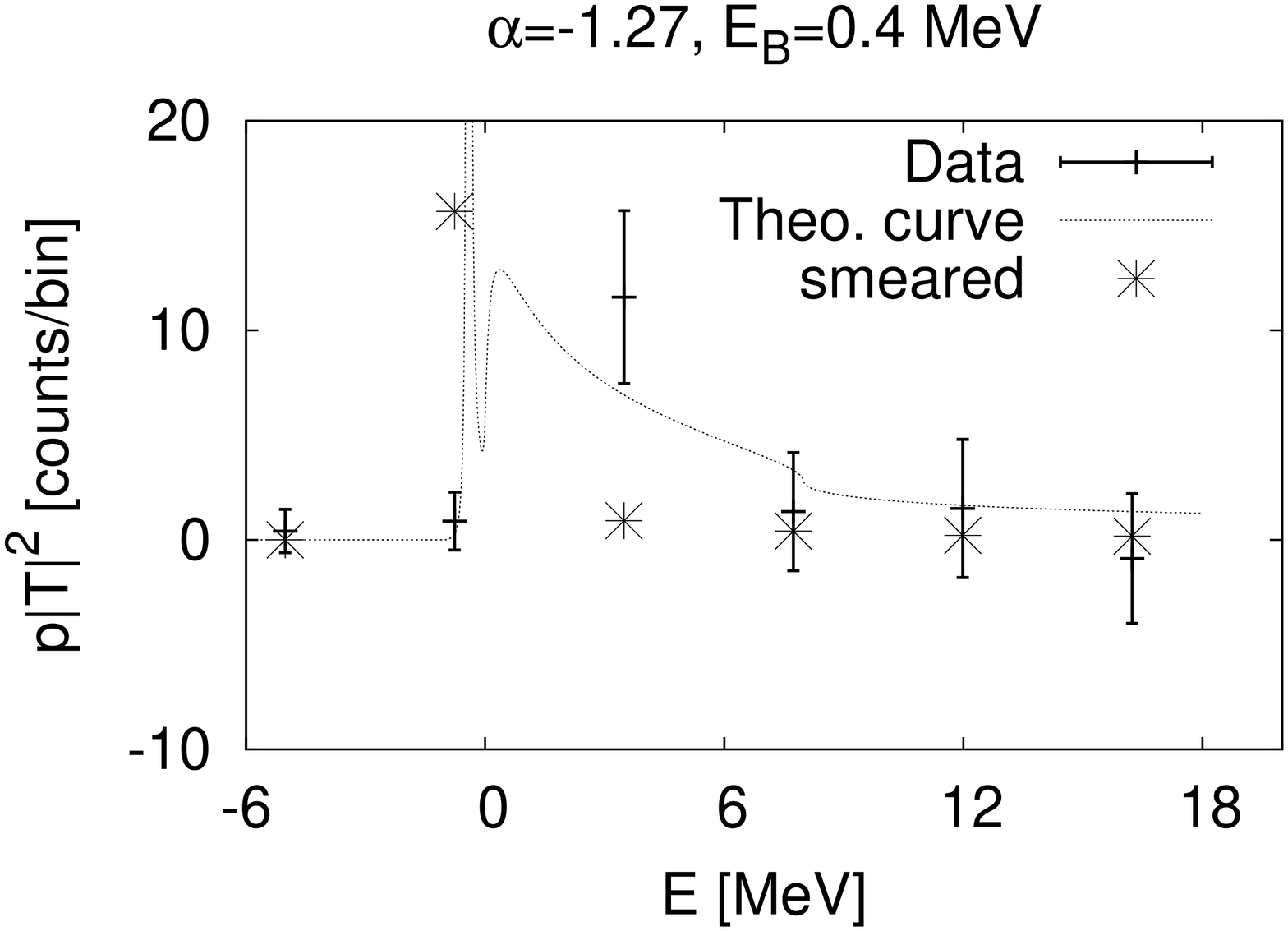} 
\end{tabular}
\caption{Theoretical results compared to data from \cite{expcross}.} \label{figcomp2}
\end{figure}
\end{widetext}

\begin{table}
\caption{$\chi^2$ values for the fits of $J/\psi \pi \pi$ and $D^0\bar D^{*0}$ production. The column to the right show the average value between the two.} \label{tabchi2}
\begin{tabular}{c||c|c|c}
\hline
 & $\chi^2$ & $\chi^2$ & \\
 $\alpha$     &    for      &   for      &   $\bar \chi^2$     \\
  & $J/\psi \pi \pi$ & $D^0\bar D^{*0}$ & \\
\hline
-1.22 & 1.83 & 0.49 & 1.16 \\
-1.23 & 1.26 & 0.50 & 0.88 \\
-1.24 & 0.87 & 0.93 & 0.90 \\
-1.25 & 0.72 & 2.77 & 1.74 \\
-1.26 & 0.92 & 17.96 & 9.44 \\
-1.27 & 0.92 & 20.31 & 10.61 \\
\hline
\end{tabular}
\end{table}


\section{Conclusion and outlook}

There is strong evidence that the $X(3872)$ state has $J^{PC}$ quantum numbers equal to $1^{++}$. It is very tempting to associate this state with a s-wave $D^0\bar D^{0*}$ molecular state if one takes into account the fact that the observed mass for this state is close to this threshold.

Using a phenomenological Lagrangian and an unitarization scheme for solving the scattering equation in couple channels, for all possible pairs of pseudoscalar and vector-mesons with zero charge and strangeness one obtains a pole with positive C-parity which can be associated with the $X(3872)$. Apart from this pole one also obtains strong attraction in the channels with negative C-parity, indicating the possible existence of a new state.

Experimentally the decays of the $X(3872)$ into $J/\psi$ with two and three pions have been measured to be of the same order of magnitude, suggesting a huge isospin violation in the decays of the $X$ state. We have studied here the effects of isospin violation in the decays of the $X$. The couplings of the $X$ to charged and neutral $D$ mesons are very similar, with at most 1.4\% of isospin violation, in our model. As a consequence of that, once one considers the decay of $X$ to $J/\psi\rho$ or $J/\psi\omega$ pairs as going through $D\bar D^*$ loops the $J/\psi\rho$ production should be suppressed in relation to the $J/\psi\omega$ by about a factor 30. If one considers the decays of the $X$ to $J/\psi$ with two and three pions as going through these channels respectively, one has also to take into account the phase-space available for each decay which, due to the $\rho$ width, is much bigger for this channel than for the $\omega$ channel, compensating the factor 30 suppressing the decays of the $X$ to $J/\psi\rho$. Moreover, the fact that the $X$ is observed in these decay modes indicates a non negligible coupling of the $X$ to vector-vector states, since it is visible in these channels ($J/\psi\rho$ and $J/\psi\omega$) even with the small phase-space available. This brought other authors to consider those channels as possible extra building blocks of the $X$ \cite{Swanson:2003tb,gutsche}, but we showed that we could explain the experimental phenomenology with dominant pseudoscalar-vector components.

The prediction of a negative C-parity state is peculiar to $D\bar D^*$ and would not appear as a vector-vector molecule. We predict this state with a framework which describes many low lying axial states and also most of the already observed axial charmed resonances. The comparison of our theoretical invariant mass distribution for $D\bar D^*$ production with data from Belle shows that one can not rule out the existence of this state and our model shows that the channels to which this resonance couples mostly are $\eta\phi$, $\eta\omega$, $\eta\prime\omega$ and $\eta_c\omega$. We also made predictions for observables where the negative C-parity state should in principle be seen and we hope these results stimulate experimental efforts in this direction.

Finally we made an investigation of the shapes of the $X(3872)$ distributions in the $J/\psi \pi \pi$ and $D^0\bar D^{*0}$ decays and found that they favor a slightly unbound, virtual state for this resonance.


\section*{Acknowledgments}

We would like to thank discussions and encouragement from E. Swanson, E. Braaten, Y. Dong and E. Lyubovitskij.

This work is 
partly supported by DGICYT contract number
FIS2006-03438. We acknowledge the support of the European Community-Research 
Infrastructure Integrating Activity
"Study of Strongly Interacting Matter" (acronym HadronPhysics2, Grant Agreement
n. 227431) under the Seventh Framework Programme of EU. 
Work supported in part by DFG (SFB/TR 16, "Subnuclear Structure of Matter").


\newpage

\newpage

\appendix

\section{The $\xi$ coefficients}

\subsection{\underline{Charge Basis}}

In charge basis we use the following ordering for the channels:

1) $\rho^+\pi^-$, 2)$K^{*+}K^-$, 3)$\rho^0\pi^0$, 4)$\omega\pi^0$, 5)$\phi\pi^0$, 6)$\rho^0\eta$, 7)$\rho^0\eta\prime$, 8)$\bar K^{*0}K^0$, 9)$D^{*+}D^-$, 10)$\bar D^{*0}D^0$, 11)$\rho^0\eta_c$, 12)$J/\psi\pi^0$, 13)$K^{*0}\bar K^0$, 14)$\rho^-\pi^+$, 15)$\omega\eta$, 16)$\phi\eta$, 17)$\omega\eta\prime$, 18)$\phi\eta\prime$, 19)$D^{*0}\bar D^0$, 20)$D^{*-}D^+$, 21)$\omega\eta_c$, 22)$\phi\eta_c$, 23)$J/\psi\eta$, 24)$J/\psi\eta\prime$, 25)$D_s^{*+}D_s^-$, 26)$D_s^{*-}D_s^+$, 27)$K^{*-}K^+$ and 28)$J/\psi\eta_c$

Tables \ref{tabcha1}, \ref{tabcha2}, \ref{tabcha3}, \ref{tabcha4} and \ref{tabcha5} show the coefficients $\xi$ for these channels.

\begin{table}[h!]
\caption{$\xi$ coefficients in charge basis basis for channels 1 to 6.} \label{tabcha1}
\begin{tabular}{c||c||c||c||c||c}
\begin{tabular}{c|c}
1 & $\xi$ \\
$\rightarrow$ & \\
\hline
1   & 2 \\
2   & 1 \\
3   & 2 \\
4   & 0 \\
5   & 0 \\
6   & 0 \\
7   & 0 \\
8   & 1 \\
9   & $\gamma$ \\
10  & $\gamma$ \\
11   & 0 \\
12  & 0 \\
13   & 0 \\
14   & 0 \\
15   & 0 \\
16   & 0 \\
17   & 0 \\
18   & 0 \\
19   & 0 \\
20  & 0 \\
21  & 0 \\
22 & 0 \\
23  & 0 \\
24  & 0 \\
25  & 0 \\
26  & 0 \\
27  & 0 \\
28  & 0 \\
\end{tabular}  &
\begin{tabular}{c|c}
2 & $\xi$ \\
$\rightarrow$ & \\
\hline
2 & 2 \\
3 &  $\frac{1}{2}$ \\
4 & $\frac{1}{2}$ \\
5 & $-\frac{1}{\sqrt{2}}$ \\
6 & $\sqrt{\frac{2}{3}}$ \\
7 & $-\frac{1}{2 \sqrt{3}}$ \\
8 & 0 \\
9 & 0 \\
10 & $\gamma$  \\
11 & 0 \\
12 & 0 \\
13 & 1 \\
14 & 0 \\
15 & $\sqrt{\frac{2}{3}}$ \\
16 & $-\frac{2}{\sqrt{3}}$ \\
17 & $-\frac{1}{2 \sqrt{3}}$ \\
18 & $\frac{1}{\sqrt{6}}$ \\
19 & 0 \\
20 & 0 \\
21 & 0 \\
22 & 0 \\
23 & 0 \\
24 & 0 \\
25 & $\gamma$  \\
26 & 0 \\
27 & 0 \\
28 & 0
\end{tabular}  &
\begin{tabular}{c|c}
3 & $\xi$ \\
$\rightarrow$ & \\
\hline
3&0 \\
4& 0 \\
5& 0 \\
6& 0 \\
7& 0 \\
8& $\frac{1}{2}$ \\
9& $\frac{\gamma }{2}$ \\
10& $\frac{\gamma }{2}$ \\
11& 0 \\
12& 0 \\
13& $\frac{1}{2}$ \\
14& 2 \\
15& 0 \\
16& 0 \\
17& 0 \\
18& 0 \\
19& $\frac{\gamma }{2}$ \\
20& $\frac{\gamma }{2}$ \\
21& 0 \\
22& 0 \\
23& 0 \\
24& 0 \\
25& 0 \\
26& 0 \\
27& $\frac{1}{2}$ \\
28& 0
\end{tabular}  &
\begin{tabular}{c|c}
4 & $\xi$ \\
$\rightarrow$ & \\
\hline
4&0 \\
5& 0 \\
6& 0 \\
7& 0 \\
8& $-\frac{1}{2}$ \\
9& $-\frac{\gamma }{2}$ \\
10& $\frac{\gamma }{2}$\\
11& 0 \\
12& 0 \\
13& $-\frac{1}{2}$ \\
14& 0 \\
15& 0 \\
16& 0 \\
17& 0 \\
18& 0 \\
19& $\frac{\gamma }{2}$ \\
20& $-\frac{\gamma }{2}$ \\
21& 0 \\
22& 0 \\
23& 0 \\
24& 0 \\
25& 0 \\
26& 0 \\
27& $\frac{1}{2}$ \\
28& 0
\end{tabular} &
\begin{tabular}{c|c}
5 & $\xi$ \\
$\rightarrow$ & \\
\hline
5& 0 \\
6& 0 \\
7& 0 \\
8& $\frac{1}{\sqrt{2}}$ \\
9& 0 \\
10& 0 \\
11& 0 \\
12& 0 \\
13& $\frac{1}{\sqrt{2}}$ \\
14& 0 \\
15& 0 \\
16& 0 \\
17& 0 \\
18& 0 \\
19& 0 \\
20& 0 \\
21& 0 \\
22& 0 \\
23& 0 \\
24& 0 \\
25& 0 \\
26& 0 \\
27& $-\frac{1}{\sqrt{2}}$ \\
28& 0
\end{tabular}  &
\begin{tabular}{c|c}
6 & $\xi$ \\
$\rightarrow$ & \\
\hline
6& 0 \\
7& 0 \\
8& $-\sqrt{\frac{2}{3}}$ \\
9& $-\frac{\gamma }{\sqrt{6}}$\\
10& $\frac{\gamma }{\sqrt{6}}$ \\
11& 0 \\
12& 0 \\
13& $-\sqrt{\frac{2}{3}} $\\
14& 0 \\
15& 0 \\
16& 0 \\
17& 0 \\
18& 0 \\
19& $\frac{\gamma }{\sqrt{6}}$ \\
20& $-\frac{\gamma }{\sqrt{6}}$ \\
21& 0 \\
22& 0 \\
23& 0 \\
24& 0 \\
25& 0 \\
26& 0 \\
27& $\sqrt{\frac{2}{3}}$ \\
28& 0
\end{tabular}
\end{tabular}
\end{table}

\begin{table}[h!]
\caption{$\xi$ coefficients in charge basis basis for channels 7 to 11.}\label{tabcha2}
\begin{tabular}{c||c||c||c||c}

\begin{tabular}{c|c}
7 & $\xi$ \\
$\rightarrow$ & \\
\hline
7& 0 \\
8& $\frac{1}{2 \sqrt{3}} $\\
9& $-\frac{\gamma }{2 \sqrt{3}}$ \\
10& $\frac{\gamma }{2 \sqrt{3}}$ \\
11& 0 \\
12& 0 \\
13& $\frac{1}{2 \sqrt{3}} $\\
14& 0 \\
15& 0 \\
16& 0 \\
17& 0 \\
18& 0 \\
19& $\frac{\gamma }{2 \sqrt{3}}$ \\
20& $-\frac{\gamma }{2 \sqrt{3}}$ \\
21& 0 \\
22& 0 \\
23& 0 \\
24& 0 \\
25& 0 \\
26& 0 \\
27& $-\frac{1}{2 \sqrt{3}}$ \\
28& 0
\end{tabular}  &

\begin{tabular}{c|c}
8 & $\xi$ \\
$\rightarrow$ & \\
\hline
8& 2 \\
9& $\gamma $ \\
10& 0 \\
11& 0 \\
12& 0 \\
13& 0 \\
14& 0 \\
15& $\sqrt{\frac{2}{3}}$ \\
16& $-\frac{2}{\sqrt{3}}$ \\
17& $-\frac{1}{2 \sqrt{3}}$ \\
18& $\frac{1}{\sqrt{6}}$ \\
19& 0 \\
20& 0 \\
21& 0 \\
22& 0 \\
23& 0 \\
24& 0 \\
25& 0 \\
26& $\gamma$  \\
27& 1 \\
28& 0
\end{tabular}  &

\begin{tabular}{c|c}
9 & $\xi$ \\
$\rightarrow$ & \\
\hline
9& $\psi +1$ \\
10& 0 \\
11& $\frac{\gamma }{\sqrt{2}} $\\
12& $\frac{\gamma }{\sqrt{2}}$ \\
13& 0 \\
14& 0 \\
15& $\frac{\gamma }{\sqrt{6}} $\\
16& 0 \\
17& $\frac{\gamma }{2 \sqrt{3}}$ \\
18& 0 \\
19& 1 \\
20& 0 \\
21& $-\frac{\gamma }{\sqrt{2}} $\\
22& 0 \\
23& $-\frac{\gamma }{\sqrt{3}}$ \\
24& $-\frac{\gamma }{\sqrt{6}}$ \\
25& 1 \\
26& 0 \\
27& 0 \\
28& $\gamma $
\end{tabular}  &

\begin{tabular}{c|c}
10 & $\xi$ \\
$\rightarrow$ & \\
\hline
10& $\psi +1 $\\
11& $-\frac{\gamma }{\sqrt{2}}$ \\
12& $-\frac{\gamma }{\sqrt{2}} $\\
13& 0 \\
14& 0 \\
15& $\frac{\gamma }{\sqrt{6}}$ \\
16& 0 \\
17& $\frac{\gamma }{2 \sqrt{3}}$ \\
18& 0 \\
19& 0 \\
20& 1 \\
21& $-\frac{\gamma }{\sqrt{2}}$ \\
22& 0 \\
23& $-\frac{\gamma }{\sqrt{3}} $\\
24&$ -\frac{\gamma }{\sqrt{6}} $\\
25& 0 \\
26& 1 \\
27& 0 \\
28& $\gamma $
\end{tabular}  &

\begin{tabular}{c|c}
11 & $\xi$ \\
$\rightarrow$ & \\
\hline
11& 0 \\
12& 0 \\
13& 0 \\
14& 0 \\
15& 0 \\
16& 0 \\
17& 0 \\
18& 0 \\
19& $-\frac{\gamma }{\sqrt{2}}$ \\
20&$ \frac{\gamma }{\sqrt{2}}$ \\
21& 0 \\
22& 0 \\
23& 0 \\
24& 0 \\
25& 0 \\
26& 0 \\
27& 0 \\
28& 0
\end{tabular} 

\end{tabular}
\end{table}

\begin{table}[h!]
\caption{$\xi$ coefficients in charge basis basis for channels 12 to 16.}\label{tabcha3}
\begin{tabular}{c||c||c||c||c}

\begin{tabular}{c|c}
12 & $\xi$ \\
$\rightarrow$ & \\
\hline
12& 0 \\
13& 0 \\
14& 0 \\
15& 0 \\
16& 0 \\
17& 0 \\
18& 0 \\
19& $-\frac{\gamma }{\sqrt{2}}$ \\
20&$ \frac{\gamma }{\sqrt{2}}$ \\
21& 0 \\
22& 0 \\
23& 0 \\
24& 0 \\
25& 0 \\
26& 0 \\
27& 0 \\
28& 0
\end{tabular} &

\begin{tabular}{c|c}
13 & $\xi$ \\
$\rightarrow$ & \\
\hline
13& 2 \\
14& 1 \\
15& $\sqrt{\frac{2}{3}}$ \\
16& $-\frac{2}{\sqrt{3}}$ \\
17& $-\frac{1}{2 \sqrt{3}}$ \\
18& $\frac{1}{\sqrt{6}}$ \\
19& 0 \\
20& $\gamma$  \\
21& 0 \\
22& 0 \\
23& 0 \\
24& 0 \\
25& $\gamma$  \\
26& 0 \\
27& 0 \\
28& 0
\end{tabular} &

\begin{tabular}{c|c}
14 & $\xi$ \\
$\rightarrow$ & \\
\hline
14& 2 \\
15& 0 \\
16& 0 \\
17& 0 \\
18& 0 \\
19& $\gamma$  \\
20& $\gamma$  \\
21& 0 \\
22& 0 \\
23& 0 \\
24& 0 \\
25& 0 \\
26& 0 \\
27& 1 \\
28& 0
\end{tabular} &

\begin{tabular}{c|c}
15 & $\xi$ \\
$\rightarrow$ & \\
\hline
15& 0 \\
16& 0 \\
17& 0 \\
18& 0 \\
19& $\frac{\gamma }{\sqrt{6}}$ \\
20& $\frac{\gamma }{\sqrt{6}} $\\
21& 0 \\
22& 0 \\
23& 0 \\
24& 0 \\
25& 0 \\
26& 0 \\
27&$ \sqrt{\frac{2}{3}}$ \\
28& 0
\end{tabular} &

\begin{tabular}{c|c}
16 & $\xi$ \\
$\rightarrow$ & \\
\hline
16& 0 \\
17& 0 \\
18& 0 \\
19& 0 \\
20& 0 \\
21& 0 \\
22& 0 \\
23& 0 \\
24& 0 \\
25& $-\frac{\gamma }{\sqrt{3}}$ \\
26& $-\frac{\gamma }{\sqrt{3}} $\\
27& $-\frac{2}{\sqrt{3}} $\\
28& 0
\end{tabular}

\end{tabular}
\end{table}

\begin{table}[h!]
\caption{$\xi$ coefficients in charge basis basis for channels 17 to 21.}\label{tabcha4}
\begin{tabular}{c||c||c||c||c}

\begin{tabular}{c|c}
17 & $\xi$ \\
$\rightarrow$ & \\
\hline
17& 0 \\
18& 0 \\
19& $\frac{\gamma }{2 \sqrt{3}}$ \\
20& $\frac{\gamma }{2 \sqrt{3}} $\\
21& 0 \\
22& 0 \\
23& 0 \\
24& 0 \\
25& 0 \\
26& 0 \\
27& $-\frac{1}{2 \sqrt{3}}$ \\
28& 0
\end{tabular} &

\begin{tabular}{c|c}
18 & $\xi$ \\
$\rightarrow$ & \\
\hline
18& 0 \\
19& 0 \\
20& 0 \\
21& 0 \\
22& 0 \\
23& 0 \\
24& 0 \\
25& $\sqrt{\frac{2}{3}} \gamma$  \\
26& $\sqrt{\frac{2}{3}} \gamma$  \\
27& $\frac{1}{\sqrt{6}}$ \\
28& 0
\end{tabular} &

\begin{tabular}{c|c}
19 & $\xi$ \\
$\rightarrow$ & \\
\hline
19& $\psi +1$ \\
20& 0 \\
21&$ -\frac{\gamma }{\sqrt{2}}$ \\
22& 0 \\
23&$ -\frac{\gamma }{\sqrt{3}}$ \\
24& $-\frac{\gamma }{\sqrt{6}}$ \\
25& 1 \\
26& 0 \\
27& $\gamma$  \\
28& $\gamma$ 
\end{tabular} &

\begin{tabular}{c|c}
20 & $\xi$ \\
$\rightarrow$ & \\
\hline
20& $\psi +1$ \\
21& $-\frac{\gamma }{\sqrt{2}}$ \\
22& 0 \\
23& $-\frac{\gamma }{\sqrt{3}}$ \\
24& $-\frac{\gamma }{\sqrt{6}} $\\
25& 0 \\
26& 1 \\
27& 0 \\
28& $\gamma $
\end{tabular} &

\begin{tabular}{c|c}
21 & $\xi$ \\
$\rightarrow$ & \\
\hline
21& 0 \\
22& 0 \\
23& 0 \\
24& 0 \\
25& 0 \\
26& 0 \\
27& 0 \\
28& 0
\end{tabular} 
\end{tabular}
\end{table}

\begin{table}[h!]
\caption{$\xi$ coefficients in charge basis basis for channels 22 to 28.}\label{tabcha5}
\begin{tabular}{c|ccccccc}
$i\rightarrow j$&22&23&24&25&26&27&28 \\
\hline
22& 0 & 0 & 0 & $-\gamma$  & $-\gamma$  & 0 & 0 \\
23& 0 & 0 & 0 & $\frac{\gamma }{\sqrt{3}}$ & $\frac{\gamma }{\sqrt{3}}$ & 0 & 0 \\
24& 0 & 0 & 0 & $-\sqrt{\frac{2}{3}} \gamma$  & $-\sqrt{\frac{2}{3}} \gamma$  & 0 & 0 \\
25& $-\gamma$  & $\frac{\gamma }{\sqrt{3}}$ & $-\sqrt{\frac{2}{3}} \gamma$  & $\psi +1$ & 0 & 0 & $\gamma$  \\
26& $-\gamma$  & $\frac{\gamma }{\sqrt{3}}$ &$ -\sqrt{\frac{2}{3}} \gamma$  & 0 &$ \psi +1 $&$ \gamma $ & $\gamma$  \\
27& 0 & 0 & 0 & 0 & $\gamma$  & 2 & 0 \\
28& 0 & 0 & 0 & $\gamma $ & $\gamma $ & 0 & 0
\end{tabular}
\end{table}


\subsection{\underline{positive C-parity basis}}

The channels here are the following:

1)$K^{*0}\bar K^0-c.c.$, 2)$\rho^+\pi^--c.c.$, 3)$D^{*0}\bar D^0-c.c.$, 4)$D^{*+}D^--c.c.$, 5)$D_s^{*+}D_s^-$ and 6)$K^{*+}K^--c.c.$

Table \ref{tabpos1} shows the $\xi$ coefficients for these channels.

\begin{table}[h!]
\caption{$\xi$ coefficients in positive C-parity basis.}\label{tabpos1}
\begin{tabular}{c|cccccc}
$i\rightarrow j$&1&2&3&4&5&6 \\
\hline
1& 2 & 1 & 0 & $\gamma$  & $-\gamma$  & -1 \\
2& 1 & 2 & $\gamma$  & $\gamma$  & 0 & 1 \\
3& 0 & $\gamma$  & $\psi +1$ & -1 & -1 & $\gamma$  \\
4& $\gamma$  & $\gamma$  & -1 & $\psi +1$ & 1 & 0 \\
5& $-\gamma$  & 0 & -1 & 1 & $\psi +1$ & $\gamma$  \\
6& -1 & 1 & $\gamma$  & 0 & $\gamma$  & 2
\end{tabular}
\end{table}

\subsection{\underline{Negative C-parity basis}}

The channels here are the following:

1) $\rho^+\pi^-+c.c.$, 2)$K^{*+}K^-+c.c.$, 3)$\rho^0\pi^0$, 4)$\omega\pi^0$, 5)$\phi\pi^0$, 6)$\rho^0\eta$, 7)$\rho^0\eta\prime$, 8)$\bar K^{*0}K^0+c.c.$, 9)$D^{*+}D^-+c.c.$, 10)$\bar D^{*0}D^0+c.c.$, 11)$\rho^0\eta_c$, 12)$J/\psi\pi^0$, 13)$\omega\eta$, 14)$\phi\eta$, 15)$\omega\eta\prime$, 16)$\phi\eta\prime$, 17)$\omega\eta_c$, 18)$\phi\eta_c$, 19)$J/\psi\eta$, 20)$J/\psi\eta\prime$, 21)$D_s^{*+}D_s^-+c.c.$ and 22)$J/\psi\eta_c$

Tables \ref{tabneg1}, \ref{tabneg2}, \ref{tabneg3} and \ref{tabneg4} show the $\xi$ coefficients for these channels.

\begin{table}[h!]
\caption{$\xi$ coefficients in negative C-parity basis basis for channels 1 to 6.} \label{tabneg1}
\begin{tabular}{c||c||c||c||c||c}

\begin{tabular}{c|c}
1 & $\xi$ \\
$\rightarrow$ & \\
\hline
1& 2 \\
2& 1 \\
3& $2 \sqrt{2}$ \\
4& 0 \\
5& 0 \\
6& 0 \\
7& 0 \\
8& 1 \\
9& $\gamma $ \\
10& $\gamma $ \\
11& 0 \\
12& 0 \\
13& 0 \\
14& 0 \\
15& 0 \\
16& 0 \\
17& 0 \\
18& 0 \\
19& 0 \\
20& 0 \\
21& 0 \\
22& 0
\end{tabular} &
\begin{tabular}{c|c}
2 & $\xi$ \\
$\rightarrow$ & \\
\hline
2& 2 \\
3& $\frac{1}{\sqrt{2}} $\\
4& $\frac{1}{\sqrt{2}}$ \\
5& -1 \\
6& $\frac{2}{\sqrt{3}}$ \\
7& $-\frac{1}{\sqrt{6}}$ \\
8& 1 \\
9& 0 \\
10& $\gamma$  \\
11& 0 \\
12& 0 \\
13&$ \frac{2}{\sqrt{3}}$ \\
14& $-2 \sqrt{\frac{2}{3}}$ \\
15& $-\frac{1}{\sqrt{6}}$ \\
16& $\frac{1}{\sqrt{3}}$ \\
17& 0 \\
18& 0 \\
19& 0 \\
20& 0 \\
21& $\gamma $ \\
22& 0
\end{tabular} &
\begin{tabular}{c|c}
3 & $\xi$ \\
$\rightarrow$ & \\
\hline
3& 0 \\
4& 0 \\
5& 0 \\
6& 0 \\
7& 0 \\
8& $\frac{1}{\sqrt{2}}$ \\
9& $\frac{\gamma }{\sqrt{2}}$ \\
10&$ \frac{\gamma }{\sqrt{2}}$ \\
11& 0 \\
12& 0 \\
13& 0 \\
14& 0 \\
15& 0 \\
16& 0 \\
17& 0 \\
18& 0 \\
19& 0 \\
20& 0 \\
21& 0 \\
22& 0
\end{tabular} &
\begin{tabular}{c|c}
4 & $\xi$ \\
$\rightarrow$ & \\
\hline
4& 0 \\
5& 0 \\
6& 0 \\
7& 0 \\
8& $-\frac{1}{\sqrt{2}}$ \\
9& $-\frac{\gamma }{\sqrt{2}}$ \\
10&$ \frac{\gamma }{\sqrt{2}}$ \\
11& 0 \\
12& 0 \\
13& 0 \\
14& 0 \\
15& 0 \\
16& 0 \\
17& 0 \\
18& 0 \\
19& 0 \\
20& 0 \\
21& 0 \\
22& 0
\end{tabular} &
\begin{tabular}{c|c}
5 & $\xi$ \\
$\rightarrow$ & \\
\hline
5& 0 \\
6& 0 \\
7& 0 \\
8& 1 \\
9& 0 \\
10& 0 \\
11& 0 \\
12& 0 \\
13& 0 \\
14& 0 \\
15& 0 \\
16& 0 \\
17& 0 \\
18& 0 \\
19& 0 \\
20& 0 \\
21& 0 \\
22& 0
\end{tabular} &
\begin{tabular}{c|c}
6 & $\xi$ \\
$\rightarrow$ & \\
\hline
6& 0 \\
7& 0 \\
8& $-\frac{2}{\sqrt{3}}$ \\
9& $-\frac{\gamma }{\sqrt{3}}$ \\
10& $\frac{\gamma }{\sqrt{3}}$ \\
11& 0 \\
12& 0 \\
13& 0 \\
14& 0 \\
15& 0 \\
16& 0 \\
17& 0 \\
18& 0 \\
19& 0 \\
20& 0 \\
21& 0 \\
22& 0
\end{tabular}

\end{tabular}
\end{table}

\begin{table}[h!]
\caption{$\xi$ coefficients in negative C-parity basis basis for channels 7 to 11.}\label{tabneg2}
\begin{tabular}{c||c||c||c||c||c}
\begin{tabular}{c|c}
7 & $\xi$ \\
$\rightarrow$ & \\
\hline
7& 0 \\
8& $\frac{1}{\sqrt{6}} $\\
9& $-\frac{\gamma }{\sqrt{6}} $\\
10& $\frac{\gamma }{\sqrt{6}}$ \\
11& 0 \\
12& 0 \\
13& 0 \\
14& 0 \\
15& 0 \\
16& 0 \\
17& 0 \\
18& 0 \\
19& 0 \\
20& 0 \\
21& 0 \\
22& 0
\end{tabular} &
\begin{tabular}{c|c}
8 & $\xi$ \\
$\rightarrow$ & \\
\hline
8& 2 \\
9& $\gamma$  \\
10& 0 \\
11& 0 \\
12& 0 \\
13& $\frac{2}{\sqrt{3}}$ \\
14& $-2 \sqrt{\frac{2}{3}}$ \\
15& $-\frac{1}{\sqrt{6}}$ \\
16& $\frac{1}{\sqrt{3}}$ \\
17& 0 \\
18& 0 \\
19& 0 \\
20& 0 \\
21& $\gamma$  \\
22& 0
\end{tabular} &
\begin{tabular}{c|c}
9 & $\xi$ \\
$\rightarrow$ & \\
\hline
9& $\psi +1$ \\
10& 1 \\
11& $\gamma$  \\
12& $\gamma $ \\
13& $\frac{\gamma }{\sqrt{3}} $\\
14& 0 \\
15& $\frac{\gamma }{\sqrt{6}}$ \\
16& 0 \\
17& $-\gamma $ \\
18& 0 \\
19& $-\sqrt{\frac{2}{3}} \gamma $ \\
20& $-\frac{\gamma }{\sqrt{3}}$ \\
21& 1 \\
22& $\sqrt{2} \gamma $
\end{tabular} &
\begin{tabular}{c|c}
10 & $\xi$ \\
$\rightarrow$ & \\
\hline
10& $\psi +1$ \\
11& $-\gamma$  \\
12& $-\gamma $ \\
13& $\frac{\gamma }{\sqrt{3}} $\\
14& 0 \\
15& $\frac{\gamma }{\sqrt{6}}$ \\
16& 0 \\
17& $-\gamma $ \\
18& 0 \\
19& $-\sqrt{\frac{2}{3}} \gamma $ \\
20& $-\frac{\gamma }{\sqrt{3}}$ \\
21& 1 \\
22& $\sqrt{2} \gamma $
\end{tabular} &
\begin{tabular}{c|c}
11 & $\xi$ \\
$\rightarrow$ & \\
\hline
11& 0 \\
12& 0 \\
13& 0 \\
14& 0 \\
15& 0 \\
16& 0 \\
17& 0 \\
18& 0 \\
19& 0 \\
20& 0 \\
21& 0 \\
22& 0
\end{tabular}
\end{tabular}
\end{table}

\begin{table}
\caption{$\xi$ coefficients in negative C-parity basis basis for channels 12 to 16.}\label{tabneg3}
\begin{tabular}{c||c||c||c||c||c}
\begin{tabular}{c|c}
12 & $\xi$ \\
$\rightarrow$ & \\
\hline
12& 0 \\
13& 0 \\
14& 0 \\
15& 0 \\
16& 0 \\
17& 0 \\
18& 0 \\
19& 0 \\
20& 0 \\
21& 0 \\
22& 0
\end{tabular} &
\begin{tabular}{c|c}
13 & $\xi$ \\
$\rightarrow$ & \\
\hline
13& 0 \\
14& 0 \\
15& 0 \\
16& 0 \\
17& 0 \\
18& 0 \\
19& 0 \\
20& 0 \\
21& 0 \\
22& 0
\end{tabular} &
\begin{tabular}{c|c}
14 & $\xi$ \\
$\rightarrow$ & \\
\hline
14& 0 \\
15& 0 \\
16& 0 \\
17& 0 \\
18& 0 \\
19& 0 \\
20& 0 \\
21& $-\sqrt{\frac{2}{3}} \gamma$  \\
22& 0
\end{tabular} &
\begin{tabular}{c|c}
15 & $\xi$ \\
$\rightarrow$ & \\
\hline
15& 0 \\
16& 0 \\
17& 0 \\
18& 0 \\
19& 0 \\
20& 0 \\
21& 0 \\
22& 0
\end{tabular} &
\begin{tabular}{c|c}
16 & $\xi$ \\
$\rightarrow$ & \\
\hline
16& 0 \\
17& 0 \\
18& 0 \\
19& 0 \\
20& 0 \\
21& $\frac{2 \gamma }{\sqrt{3}}$ \\
22& 0
\end{tabular}
\end{tabular}
\end{table}

\begin{table}
\caption{$\xi$ coefficients in negative C-parity basis basis for channels 17 to 22.}\label{tabneg4}
\begin{tabular}{c|cccccc}
$i\rightarrow j$&17&18&19&20&21&22 \\
\hline
17& 0 & 0 & 0 & 0 & 0 & 0 \\
18& 0 & 0 & 0 & 0 & $-\sqrt{2} \gamma$  & 0 \\
19& 0 & 0 & 0 & 0 & $\sqrt{\frac{2}{3}} \gamma $ & 0 \\
20& 0 & 0 & 0 & 0 & $-\frac{2 \gamma }{\sqrt{3}}$ & 0 \\
21& 0 & $-\sqrt{2} \gamma $ & $\sqrt{\frac{2}{3}} \gamma$  & $-\frac{2 \gamma }{\sqrt{3}}$ & $\psi +1 $& $\sqrt{2} \gamma $ \\
22& 0 & 0 & 0 & 0 &$ \sqrt{2} \gamma $ & 0
\end{tabular}
\end{table}

\newpage

\newpage

\hbox{        }


\newpage

\newpage


\begin{thebibliography}{99}

\bibitem{belledisc}
  S.~K.~Choi {\it et al.}  [Belle Collaboration],
  Phys.\ Rev.\ Lett.\  {\bf 91}, 262001 (2003)
  [arXiv:hep-ex/0309032].

\bibitem{conf1}
  D.~E.~Acosta {\it et al.}  [CDF II Collaboration],
  Phys.\ Rev.\ Lett.\  {\bf 93}, 072001 (2004)
  [arXiv:hep-ex/0312021].

\bibitem{conf2}
  V.~M.~Abazov {\it et al.}  [D0 Collaboration],
  Phys.\ Rev.\ Lett.\  {\bf 93}, 162002 (2004)
  [arXiv:hep-ex/0405004].

\bibitem{conf3}
  B.~Aubert {\it et al.}  [BABAR Collaboration],
  Phys.\ Rev.\  D {\bf 71}, 071103 (2005)
  [arXiv:hep-ex/0406022].


\bibitem{cdf2pi}
  A.~Abulencia {\it et al.}  [CDF Collaboration],
  Phys.\ Rev.\ Lett.\  {\bf 96}, 102002 (2006)
  [arXiv:hep-ex/0512074].

\bibitem{bellegj}
  K.~Abe {\it et al.},
  arXiv:hep-ex/0505037.
  
\bibitem{xqn}
  A.~Abulencia {\it et al.}  [CDF Collaboration],
  Phys.\ Rev.\ Lett.\  {\bf 98}, 132002 (2007)
  [arXiv:hep-ex/0612053].
  
\bibitem{nojeta}
  B.~Aubert {\it et al.}  [BABAR Collaboration],
  Phys.\ Rev.\ Lett.\  {\bf 93}, 041801 (2004)
  [arXiv:hep-ex/0402025].
  
\bibitem{Terasaki:2007uv}
  K.~Terasaki,
  arXiv:0706.3944 [hep-ph].
  
\bibitem{danielaxial}
  D.~Gamermann and E.~Oset,
  Eur.\ Phys.\ J.\  A {\bf 33}, 119 (2007)
  [arXiv:0704.2314 [hep-ph]].
  
\bibitem{Liu:2008fh}
  Y.~R.~Liu, X.~Liu, W.~Z.~Deng and S.~L.~Zhu,
  Eur.\ Phys.\ J.\  C {\bf 56}, 63 (2008)
  [arXiv:0801.3540 [hep-ph]].
  
\bibitem{Liu:2007bf}
  X.~Liu, Y.~R.~Liu, W.~Z.~Deng and S.~L.~Zhu,
  Phys.\ Rev.\  D {\bf 77}, 034003 (2008)
  [arXiv:0711.0494 [hep-ph]].
  
\bibitem{Dong:2008gb}
  Y.~b.~Dong, A.~Faessler, T.~Gutsche and V.~E.~Lyubovitskij,
  Phys.\ Rev.\  D {\bf 77}, 094013 (2008)
  [arXiv:0802.3610 [hep-ph]].
  
\bibitem{Swanson:2003tb}
  E.~S.~Swanson,
  Phys.\ Lett.\  B {\bf 588}, 189 (2004)
  [arXiv:hep-ph/0311229].

\bibitem{gutsch1}
  M.~B.~Voloshin,
  Phys.\ Lett.\  B {\bf 604}, 69 (2004)
  [arXiv:hep-ph/0408321].

\bibitem{gutsch2}
  E.~Braaten and M.~Kusunoki,
  Phys.\ Rev.\  D {\bf 72}, 054022 (2005)
  [arXiv:hep-ph/0507163].
  
\bibitem{Liu:2008du}
  X.~Liu, Y.~R.~Liu and W.~Z.~Deng,
  arXiv:0802.3157 [hep-ph].

\bibitem{gutsche}
  Y.~Dong, A.~Faessler, T.~Gutsche, S.~Kovalenko and V.~E.~Lyubovitskij,
  arXiv:0903.5416 [].
  
 \bibitem{braaten} E. Braaten, talk at the International Workshop on 
 Effective Field Theories: from the Pion to the Upsilon.
 http://ific.uv.es/eft09/

\bibitem{qq1}
  E.~S.~Swanson,
  Phys.\ Rept.\  {\bf 429}, 243 (2006)
  [arXiv:hep-ph/0601110].

\bibitem{qq2}
  G.~Bauer,
  Int.\ J.\ Mod.\ Phys.\  A {\bf 21}, 959 (2006)
  [arXiv:hep-ex/0505083].

\bibitem{qq3}
  M.~B.~Voloshin,
  Prog.\ Part.\ Nucl.\ Phys.\  {\bf 61}, 455 (2008)
  [arXiv:0711.4556 [hep-ph]].

\bibitem{olsen1}
  S.~L.~Olsen,
  arXiv:0901.2371 [hep-ex].

\bibitem{expcross}
 G.Majumder, ICHEP2006 talk, http://belle.kek.jp/belle/talks/ICHEP2006/Majumber.ppt.

\bibitem{hyodo}
  T.~Hyodo, D.~Jido and A.~Hosaka,
  Phys.\ Rev.\  C {\bf 78}, 025203 (2008)
  [arXiv:0803.2550 [nucl-th]].

\bibitem{ollerloop}
 J.~A.~Oller and U.~G.~Meissner,
 Phys.\ Lett.\  B {\bf 500}, 263 (2001)
 [arXiv:hep-ph/0011146].


\bibitem{jnieves}
  H.~Toki, C.~Garcia-Recio and J.~Nieves,
  Phys.\ Rev.\  D {\bf 77}, 034001 (2008)
  [arXiv:0711.3536 [hep-ph]].

\bibitem{hana0}
  V.~Baru, J.~Haidenbauer, C.~Hanhart, Yu.~Kalashnikova and A.~E.~Kudryavtsev,
  Phys.\ Lett.\  B {\bf 586} (2004) 53
  [arXiv:hep-ph/0308129].

\bibitem{weinberg}
  S.~Weinberg,
  Phys.\ Rev.\  {\bf 130}, 776 (1963).

\bibitem{efimov}
 G.~V.~Efimov and M.~A.~Ivanov,
 IOP Publishing, Bristol \& Philadelphia (1993).

\bibitem{hidden1}
  H.~Nagahiro, L.~Roca, A.~Hosaka and E.~Oset,
  Phys.\ Rev.\  D {\bf 79}, 014015 (2009)
  [arXiv:0809.0943 [hep-ph]].

\bibitem{hidden2}
  L.~S.~Geng and E.~Oset,
  arXiv:0812.1199 [hep-ph].

\bibitem{raquelhidden}
  R.~Molina, H.~Nagahiro, A.~Hosaka and E.~Oset,
  arXiv:0903.3823 [hep-ph].

\bibitem{withzou}
  D.~Gamermann, E.~Oset and B.~S.~Zou,
  arXiv:0805.0499 [hep-ph].

\bibitem{hanhart}
  C.~Hanhart, Yu.~S.~Kalashnikova, A.~E.~Kudryavtsev and A.~V.~Nefediev,
  Phys.\ Rev.\  D {\bf 76}, 034007 (2007)
  [arXiv:0704.0605 [hep-ph]].

\bibitem{braaten2}
  E.~Braaten and M.~Lu,
  Phys.\ Rev.\  D {\bf 76}, 094028 (2007)
  [arXiv:0709.2697 [hep-ph]].

\bibitem{voloshin2}
  M.~B.~Voloshin,
  Phys.\ Lett.\  B {\bf 579}, 316 (2004)
  [arXiv:hep-ph/0309307].

\bibitem{braaten3}
  E.~Braaten and M.~Lu,
  Phys.\ Rev.\  D {\bf 77}, 014029 (2008)
  [arXiv:0710.5482 [hep-ph]].





\end{thebibliography}
\end{document}